\documentstyle[12pt,overcite]{article}

\input epsf

\def\ba{\begin{eqnarray}}
\def\ea{\end{eqnarray}}
\def\be{\begin{equation}}
\def\ee{\end{equation}}

\def\mxth{\mathsurround=0pt }
\def\xversim#1#2{\lower2.pt\vbox{\baselineskip0pt \lineskip-.2pt
    \ialign{$\mxth#1\hfil##\hfil$\crcr#2\crcr\sim\crcr}}}

    \def\lsim{\mathrel{\mathpalette\xversim <}}
    \def\gsim{\mathrel{\mathpalette\xversim >}}

\newcommand{\labeq}[1] {\label{eq:#1}}

\begin{document}


\begin{center}
\Large{Cosmic Evolution in a Cyclic Universe}
\end{center}
\vspace{.1in}

\begin{center}
Paul J. Steinhardt$^1$ and Neil Turok$^{2}$ \\
{\it
$^1$Joseph Henry Laboratories,
Princeton University,
Princeton, NJ 08544, USA \\
$^2$DAMTP, CMS, Wilberforce Road, Cambridge, CB3 0WA, UK}
\end{center}

\vspace{.5in}
\noindent
Based on concepts drawn from the ekpyrotic scenario and M-theory, 
we elaborate our  recent proposal 
of a cyclic model of the Universe.
In this model, the Universe undergoes an endless sequence 
of cosmic epochs
 which begin with the Universe expanding from a `big bang'
and end with the Universe contracting to a `big crunch.'  
 Matching from `big crunch' to
`big bang' is performed according to the prescription 
recently proposed with Khoury, Ovrut and Seiberg.
The expansion part of 
the cycle includes a period of radiation and  matter domination
followed by an extended period of cosmic acceleration at low energies.
The cosmic acceleration
is crucial in establishing the flat and vacuous initial conditions 
required for ekpyrosis and for removing the entropy,
black holes, and other debris produced in the preceding cycle.
By restoring the Universe to the same  vacuum state before each big 
crunch, the acceleration insures that the cycle can repeat 
and that the cyclic solution is an attractor.

\vspace*{.1in}

\noindent
PACS number(s):
11.25.-w,04.50.+h, 98.80.Cq,98.80.-k
%

\newpage

\section{Introduction}

In a recent paper,\cite{ST1}
we introduced the possibility of a cyclic Universe,
a cosmology in which the Universe undergoes a periodic sequence
of expansion and contraction.  Each cycle
begins with a ``big bang" and ends in a ``big crunch," 
only to emerge in a big bang once again.
The expansion phase of each cycle includes a period of radiation-,
matter-, and quintessence-domination, the last phase of which corresponds
to the current epoch of cosmic acceleration. The accelerated expansion
phase dilutes by an exponential factor the entropy 
and the density of black holes and any other debris
produced since  the preceding big bang. 
The acceleration 
 ultimately ends, and it is followed by a
 period of decelerating expansion  and then contraction.
At the transition from big crunch to big bang, matter and radiation
are created, restoring the Universe to the high density required
for a new  big bang phase.

Historically, cyclic models have been considered attractive because
they avoid the issue of initial conditions.\cite{Rocher}  Examples can be
found in mythologies and philosophies dating back  to the beginning of 
recorded 
history.  Since the introduction of general relativity, though,
various problems with the
cyclic concept  have emerged. In the 1930's,
Richard Tolman\cite{Tolman}
discussed cyclic models 
consisting of a closed Universe with zero cosmological constant.
He pointed out that entropy generated in one cycle would add
to the entropy created in the next.   Consequently,
the maximal size of the Universe, and the duration of
a cycle, increase from bounce to bounce.
Extrapolating backwards, the  duration of the bounce converges
to zero in a finite time. Consequently, the problem of initial
conditions remains.  In the 1960's, the singularity 
theorems of Hawking and Penrose showed that a big crunch necessarily
leads to a cosmic singularity where general relativity 
becomes invalid.  Without a theory to replace general relativity
in hand, considerations of whether time and space could exist 
before the big bang were discouraged.  ``Big bang" became 
synonymous with the  beginning of space-time.  However,
there is nothing in the Hawking-Penrose singularity theorems
to suggest that cyclic behavior is
forbidden in an improved theory of gravity, such as string theory
and M theory, 
and some people have 
continued to speculate on this possibility.\cite{DP,PBB}
In the 1990's, observations showed that
the matter density is significantly
less than the critical density and that the scale factor of the 
Universe is accelerating.\cite{rev}   Tolman's cyclic model based on a
closed Universe is therefore observationally ruled out.

Curiously, the same observations that eliminate Tolman's cyclic
model fit perfectly the novel kind of cyclic model proposed here.
In our proposal,
the Universe is flat, rather than closed. 
The transition from expansion to contraction is caused by 
introducing negative potential energy,  rather than 
spatial curvature.  Furthermore, the cyclic behavior depends 
in an essential way 
on having a period of accelerated expansion {\it after} the 
radiation and matter-dominated phases.  During
the accelerated expansion phase, 
the Universe approaches a  nearly vacuous state, 
restoring very nearly identical local conditions 
as existed in the previous 
cycle prior to the contraction phase.  
Globally, 
the total entropy in the Universe grows from cycle to 
cycle, as Tolman suggested.  However, the entropy density,
which is all any real observer would actually see, 
has perfect cyclic behavior with entropy density 
being created at each bounce, and subsequently being 
diluted to negligible levels before the next bounce. 

The linchpin of the cyclic picture is safe passage through
the cosmic singularity, the transition from 
the big crunch to big bang.  
In recent work with J. Khoury, B. Ovrut and N. Seiberg, we 
have proposed that a smooth  transition is possible
in string theory.\cite{kost1,nonsing}
In ordinary 4d general relativity,
the big crunch is  interpreted
as the collapse and disappearance of four-dimensional space-time.
Densities and curvatures diverge and
and there is no sign that a transition is possible.
But in the theory considered here,  what appears to be 
a big crunch in the 4d effective theory actually 
corresponds to the momentary
collapse of an additional fifth dimension. As far as matter which
couples to the higher dimensional metric is concerned, 
the three large spatial dimensions remain large and time
continues smoothly.  
The temperature and density are
finite as one approaches the crunch, and, furthermore,  the 
geometry is flat just before and just after the bounce.
In short,
there is nothing to suggest that time comes to an end when 
the fifth spatial dimension collapses.
Quite the contrary, the most natural possibility is that time
continues smoothly.
Efforts are currently
underway to establish this conclusion rigorously in string 
theory.\cite{SP}
The cyclic scenario considered here exploits this concept and 
is absolutely dependent on its validity.
In the absence of a detailed theory of the transition from
big crunch to big bang,
we will parameterize the bounce in terms of simple matching 
conditions incorporating energy and momentum conservation.

The appeal of a cyclic model is that it
provides a description of the 
history of the Universe which applies
arbitrarily far back into our past. The model 
presented here suggests novel answers to some of the most 
challenging issues in cosmology: How old is the Universe -
finite or infinite? How large is it?
What was the physical cause of its homogeneity, isotropy and flatness?
What was
the origin of the energy density inhomogeneities that seeded cosmic
structure formation and are visible on the cosmic 
microwave sky?  What is the resolution of the cosmic singularity puzzle?
Was there time, and an arrow of time, before the big bang?
In addition, 
our scenario has a number of  surprising implications for other
major puzzles such as the value 
of the cosmological constant, the relative densities of different
forms of matter,  and even for
supersymmetry breaking.

The cyclic  model 
rests heavily on  ideas
developed as part of   the recently proposed 
``ekpyrotic Universe."\cite{kost1,nonsing}
The basic  physical notion is that 
the collision between two brane worlds approaching one
another along an
extra dimension would have literally
generated a hot big bang.
Although the original ekpyrosis
paper focused on
collisions between bulk branes and boundary branes,\cite{kost1}
here the
more relevant example is where the boundary branes collide,
the extra dimension disappears momentarily  and the 
branes then bounce apart.\cite{nonsing}
The ekpyrotic scenario introduced several important  concepts that
serve as building blocks for the cyclic scenario:
\begin{itemize}
\item boundary branes approaching one another (beginning from rest)
corresponds to contraction in the 
effective 4d theoretic description;\cite{kost1}
\item contraction produces a  blue shift effect that converts
gravitational energy into brane kinetic energy;\cite{kost1}
\item collision converts some fraction of brane kinetic energy 
into matter and radiation that can fuel the big bang;\cite{kost1,ekperts}
\item the collision and bouncing apart of boundary branes corresponds
to the transition from a big crunch to a big bang.\cite{nonsing}
\end{itemize}

A  key element is added to obtain a cyclic Universe.
The ekpyrotic scenario assumes that there is only one collision 
after which the interbrane potential becomes zero
(perhaps due to changes in the gauge degrees of freedom on the
branes that zero out the force).
The cyclic model assumes instead that the interbrane potential is the 
same before and after collision.
After  the branes  bounce and fly apart, the  interbrane potential 
ultimately causes them to draw together  and collide again.
To ensure cyclic behavior, we will show that the potential must 
vary from negative to positive values.
(In  the ekpyrotic
examples, the potentials are zero or negative for 
all inter-brane separations.)
We propose that, 
at distances corresponding to the present-day 
separation between the branes,
the inter-brane potential energy density should be positive and
correspond to the currently observed  
dark energy, providing roughly 70\% of the  critical density today.
That is, the dark energy that is causing the cosmic
acceleration of the Universe today is, in this scenario, 
inter-brane potential energy.
The 
dark energy and its 
associated cosmic acceleration play an essential role 
in restoring the Universe to a nearly vacuous state
thereby allowing the cyclic solution to become
an attractor. 
As the brane separation decreases, 
the interbrane potential becomes negative, 
as in the ekpyrotic scenario.
As the branes approach one another,
the scale factor of the Universe, in the conventional
Einstein description, changes from expansion to contraction.  When the branes
collide and bounce, matter and radiation are produced and
there is a second reversal transforming
contraction to expansion so a new cycle can begin.

The central element in the cyclic scenario is
a four dimensional scalar field $\phi$, 
parameterizing the inter-brane 
distance or equivalently the size of the fifth dimension. The
brane separation goes to zero as $\phi$ 
tends to $-\infty$, and the maximum brane separation 
is attained at some finite value $\phi_{max}$.
For the most part our discussion will be framed
completely within the four dimensional
effective theory of gravity and matter coupled to
the scalar field $\phi$. 
This description is universal in the sense
that 
many higher dimensional brane models 
converge to the same
four dimensional effective description in the limit of small 
brane separation.
We shall not need to tie ourselves to a particular
realization of the brane world idea, such as heterotic
M theory for the purposes of this discussion, 
although such an underlying description 
is certainly required, 
both for actually deriving the scalar potential
we shall simply postulate and 
for the ultimate quantum consistency of the theory. 
The extra dimensional, and string theoretic interpretation, is
also  crucial at  
the brane collision, where the effective four
dimensional Einstein-frame description is singular and 
at which point we postulate a big crunch-big bang transition as outlined in 
Ref.~(\ref{nonsing}). Again, for the present discussion we
simply
parameterize the outcome of this transition in terms
of the density of radiation produced on the branes, and 
the change in the kinetic energy of the scalar field,
corresponding to a change in the 
contraction/expansion rate of the fifth dimension.

The scalar field $\phi$ plays a crucial role in the cyclic
scenario, in regularizing
the Einstein-frame singularity. Matter and
radiation on the brane 
couple to the Einstein frame scale factor $a$ times
a function $\beta(\phi)$ with exponential
behavior as $\phi \rightarrow -\infty$, such that the product
is generically finite at the brane collision, even though
$a=0$ and $\phi=-\infty$ there. For finite $\phi$,
the coupling of the matter and
radiation to $\phi$ is more 
model-dependent. Models in which $\phi$ is massless 
at the current epoch, such as we describe in this paper, 
face a strong constraint due to the fact that
$\phi$ can mediate a `fifth force', which is
in general composition dependent and violates
the equivalence principle. Again, without
tying ourselves to a particular brane world scenario
we shall consider models in which the coupling function
$\beta(\phi)$ tends to a constant at current values of
$\phi$ (large brane separations),
and the corresponding fifth force is weak. An example
of such a model is the Randall-Sundrum model with 
the non-relativistic matter we are made of
localized on the positive tension 
brane (see e.g. Ref.~\ref{PGT} for a recent discussion). 
In models where $\beta(\phi)$ does not tend to
a constant at current values of $\phi$, one must invoke some
physical mechanism to give
the $\phi$ field 
a small mass so that the fifth force is only 
short-ranged. This modification still allows for
cyclic behavior, with an epoch of false vacuum domination
followed by tunneling 
\cite{SST}.

The outline of this paper is as follows.
In Section~\ref{sec2}, we describe the requisite properties of the
scalar field (inter-brane) potential  and
present a brief tour through one complete cosmic cycle.
In subsequent sections, we focus in technical detail  on
various stages of the cycle: 
the bounce (Section~\ref{sec3}), passing through the potential well after
the big bang (Section~\ref{sec4}),
the radiation-, matter- and quintessence- dominated epochs (Section~\ref{sec5}),
 the onset of the contraction phase and the generation of    
  density perturbations (Section~\ref{sec6}).
 In Section~\ref{sec7}, we show that the cyclic solution is a stable
 attractor solution under classical and quantum fluctuations.
In Section~\ref{sec8}, we discuss the implications for the fundamental
questions of cosmology  introduced above.

\section{A Brief Tour of the Cyclic Universe} \label{sec2}

The various stages of a cyclic model can be characterized in terms of
a scalar field $\phi$  which moves back and forth in an
effective potential $V(\phi)$. In Section~\ref{sec2a}, we discuss
the basic properties that $V(\phi)$ must have in order to allow
cyclic solutions. 

The stages of expansion and contraction can be described from two
different points of view.  First, one can choose fields and coordinates
so that the full  extra-dimensional theory is reduced to an effective
four-dimensional theory with a conventional Einstein action.  
The key parameters are the scale factor $a$ and the modulus
scalar field $\phi$ that determines the distance between branes.
In this picture, the terms ``big bang" and ``big crunch" seem 
well-merited.  The scale factor collapses to zero at the big crunch,
bounces, and grows again after the big bang. However, what is 
novel is the presence of the scalar field $\phi$  which is runs to
$-\infty$ at the bounce with diverging kinetic energy.  The 
scalar field acts as a fifth force, modifying in an essential way
the behavior of matter and energy at the big crunch. Namely, 
the temperature and matter density remain finite at the bounce
because the usual blue shift effect during contraction is compensated
by the fifth force effect due to $\phi$. The arrangement seems 
rather magical if one is unaware that the 4d theory is derived from
a higher dimensional picture in which this behavior has a clear
geometrical interpretation. Nevertheless,
for most of this paper
we shall keep to the four dimensional Einstein description,
switching to the higher dimensional picture only when 
necessary to understand the bounce, or to discuss global
issues where matching one cycle to the next is important. 
The description of a single cycle from the 4d effective
theory point-of-view is given in Section~\ref{sec2b}.

The same evolution appears to be quite different to observers
on the visible brane who detect matter and radiation confined to 
three spatial dimensions.  In this picture, depending on the details,
the brane is either always, or nearly always expanding
except for tiny jags near the big crunch/big bang transition
when it contracts by 
a modest amount.
The branes stretch at a rate that depends on which form of energy 
dominates the energy density of the Universe.  As the big crunch is 
approached, however, the expansion rate changes suddenly, and 
new matter and radiation is created (a brane has instantaneously
collided into the visible brane and bounced from it).
We describe some aspects of the visible brane viewpoint 
in Section~\ref{sec2c}.

This picture makes it clear that the big crunch does not correspond to 
the disappearance of all of space and the end of time 
but, rather, to the momentary disappearance of a fifth dimension.
However, the behavior of gravity itself appears quite wild because
it depends on the full bulk space-time, which is changing rapidly.
One way of describing this picture is that one has mapped the
conventional big bang singularity onto the mildest form of
singularity possible, namely the disappearance of a single
dimension for an instant of time. Nevertheless there are delicate
issues involved, as are discussed in Ref.~\ref{nonsing}, 
such as the fact that the effective four dimensional Planck 
mass hits zero at the singularity, so that gravitational
fluctuations can become large. There are suggestions in specific
calculations\cite{ekperts} that physical  quantities are
nevertheless well behaved although a great deal more remains 
to be done
to make the picture rigorous.

\subsection{The Effective Potential for a Cyclic Universe}\label{sec2a}

We will consider in this
paper potentials  $V(\phi)$ of the form
 shown  in Figure~1, with the following key
 features:

$\bullet$ The potential tends to zero rapidly as
$\phi \rightarrow -\infty$. One natural possibility for
the extra dimension parameterized by $\phi$ is
the eleventh dimension of M theory. In this case the string
coupling constant $g_s \propto e^{\gamma \phi}$, with
some positive constant $\gamma$, and $g_s$ 
vanishes as $\phi \rightarrow -\infty$.
Non-perturbative potentials should vanish faster than
any finite power of $g_s$, {\it  i.e.,} faster than an exponential
in $\phi$.

$\bullet$ The potential is negative for intermediate $\phi$, 
and rises with a region of large negative curvature, $V''/V >>1$
covering a range of $\phi$ of order unity in Planck mass units.
This region is required for the production of scale invariant
density perturbations, as proposed in Ref.~\ref{kost1}
and detailed in Ref.~\ref{ekperts}.
Attractive exponential
potentials of this  type could be produced, for example, by
the virtual 
exchange of massive particles between the boundary branes. 

$\bullet$ As $\phi$ increases, the potential rises to
a shallow plateau, with $V''/V <<1$ and a {\it positive}
height $V_0$ given
by the present vacuum energy of the Universe as inferred from
cosmic acceleration and other astronomical evidence. 
The positive energy density is essential for having a cyclic
solution since it produces a period of cosmic acceleration that
restores the Universe to a nearly vacuous state before the next bounce.
The discussion
here can be extended to potentials of a more general form.  
For example,
it is not essential that the  positive plateau persist to arbitrarily
large $\phi$
since the cyclic solution only explores a finite range of $\phi > 0$. 
Provided  the 
condition $V''/V \ll 1$ is satisfied over that range,
 the Universe undergoes cosmic acceleration
when the field rolls down that portion of the potential.
However, for simplicity, we will consider the example in Figure~1.

An explicit model for $V(\phi)$ which is convenient
for analysis is
\be
V(\phi)= V_0(1 - e^{-c \phi}) F(\phi),
\labeq{examplep}
\ee
where, without loss of generality,
we have shifted $\phi$ so that the zero of the potential
occurs at $\phi=0$. 
The function $F(\phi)$  is introduced to
represent the vanishing of non-perturbative effects
described above: $F(\phi)$
turns off the potential rapidly as $\phi$ goes below 
$\phi_{min}$, but it approaches one for $\phi > \phi_{min}$.
For example, $F(\phi)$  might be proportional 
to $e^{-1/g_s^2} $   or $e^{-1/g_s}$,
where $g_s \propto e^{\gamma \phi}$ for $\gamma > 0$.
The constant $V_0$ is set roughly equal to
the vacuum energy observed in today's Universe, of order 
$10^{-120}$ in Planck units. We do not attempt to explain this number.
Various suggestions as to how a suitable small positive
vacuum energy could arise have been made.cite{sw,bp} For large $c$,
this potential has $V''/V \ll 1$ for $\phi \gsim 1$ and $V''/V \gg 1$
for $\phi_{min} < \phi<0$.  These two regions  account
for cosmic acceleration and for ekpyrotic production of
density perturbations,  respectively.\cite{kost1,ekperts} 
In the latter  region, the constant term is irrelevant and
$V$ may be approximated by
$ -V_0 \, e^{-c \phi}$ which may be studied 
using the scaling solution discussed 
in Section~\ref{sec6}.

For an arbitrary scalar potential of the form sketched, i.e.
rising with negative curvature towards a flat plateau, the
the scalar
spectral index 
is given approximately by\cite{ekperts,fut}
\be \label{genn}
n_S \approx 1 - 4\left[ 1+  \left(\frac{V}{V'}\right)^2 
 - \frac{V''V}{(V')^2}\right],
\ee 
to be evaluated when the modes on the length scales 
of interest are generated, Stage~(4)
as described in Fig.~1.
For the exponential form here, Eq.~(\ref{genn}) reduces to
\be
n_S \approx 1-{4\over c^2}.
\labeq{spec}
\ee
Current observational limits from
the cosmic microwave background
and large scale structure data are safely satisfied
for  $c=10$, which we shall adopt as our
canonical value.

\begin{figure} \label{potent}
\epsfxsize=5 in \centerline{\epsfbox{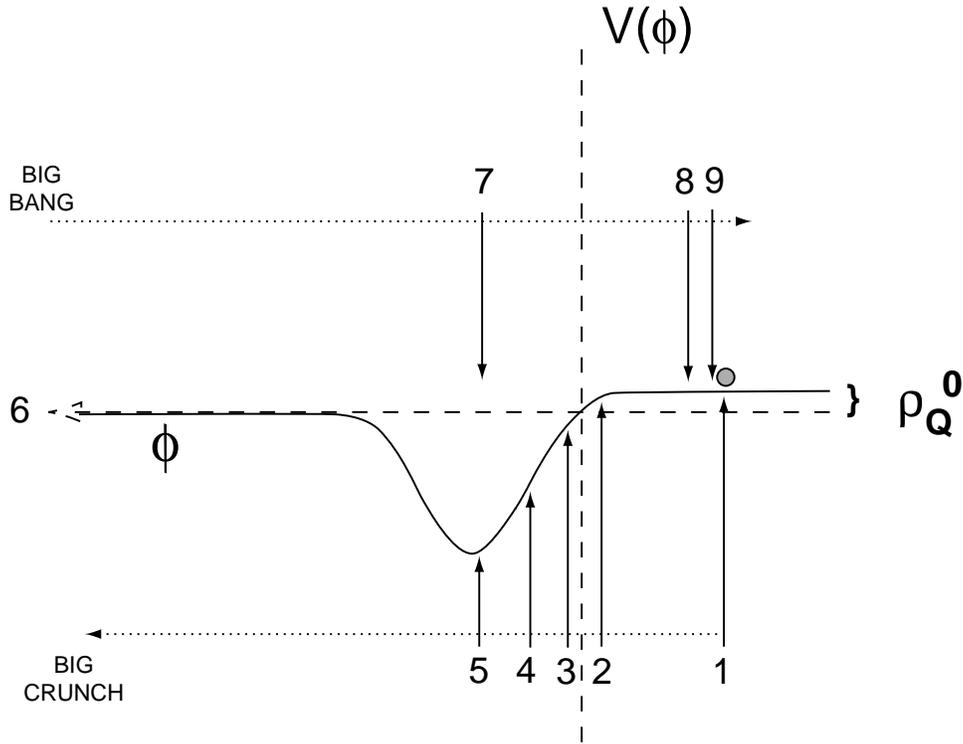}}
 \caption{
The interbrane potential $V(\phi)$ versus $\phi$, whose 
value ($-\infty<\phi<\phi_{\infty}$) determines the distance between branes.
The shaded circle represents the maximum positive value
of $\phi$ during the cycle. The various
stages are: 
(1) quintessence/potential domination and 
cosmic acceleration (duration~$\ge$ trillion years); 
(2) 
$\phi$ kinetic
energy becomes non-negligible, decelerated expansion 
begins (duration~$\sim$ 1 billion years); 
(3) $H=0$, contraction begins; (4) 
density fluctuations on observed scales created ($(t_0 t_R)^{1/2} 
\approx 1$~ms before big crunch);  (5) $\phi$ 
kinetic energy domination begins ($t_{min} \sim
10^{-30}$~s before big crunch); (6) bounce and reversal from 
big crunch to big bang; (7) end of $\phi$ kinetic energy domination,
potential also contributes
($t_{min} \sim
10^{-30}$~s after big bang); (8) radiation dominated epoch begins
$t_R \sim 10^{-25}$~s after big bang);
(9) matter domination epoch begins ($\sim 10^{10}$~s after
big bang). 
As the potential begins
to dominate and the Universe  returns to stage (1),
the field turns around and rolls back towards $-\infty$.
}
 \end{figure}

The fact that
the potential minimum is negative means that there are no strictly
static solutions for $\phi$ except 
anti-de Sitter space.  However, as we shall show, the generic
behavior --  indeed an attractor --  is a dynamical ``hovering"
solution in which 
$\phi$ roams back and forth in cyclic fashion between the
plateau and $-\infty$.  The hovering solution is highly asymmetric 
in time.  The field $\phi$ spends trillions of years or more
on the plateau and mere instants traveling from the potential
well to $-\infty$ and back.   Gravity and the bounce provide
transfers of gravitational to kinetic to matter-radiation density
that keep the Universe forever hovering around the anti-de Sitter
minimum rather than being trapped in it.

\subsection{The View from Effective 4d Theory} \label{sec2b}

\begin{figure} \label{avst}
 \epsfxsize=3.3 in \centerline{\epsfbox{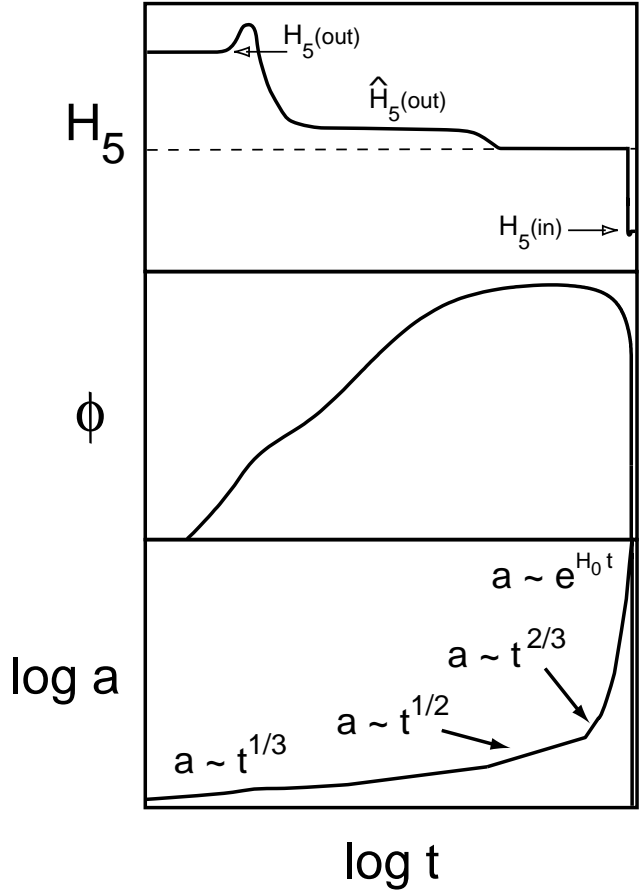}}
    \caption{
    Schematic plot of the scale factor $a(t)$,  the modulus $\phi(t)$,
    and $H_5 \equiv {2\over 3} d( {\rm exp}(\sqrt{3/2} \phi)/dt $
    for one cycle, where $t$ is Einstein frame proper time.
    The scale factor starts out zero but
    expands as $t^{1\over 3}$,
    and the scalar field grows logarithmically with $t$, 
    in the scalar kinetic energy
    dominated early regime. Then, when radiation begins to dominate
    we have $a \propto t^{1\over 2}$, and the scalar field motion is
    strongly damped. This is followed by the
    matter era, where $a \propto t^{2\over 3}$, and a 
    potential
    dominated phase in which $a(t)$ increases exponentially, before a final
    collapse on a timescale $H_0^{-1}$, to $a=0$ once more.
    $H_5$ is proportional to
    the proper (five dimensional) 
    speed of contraction of the fifth dimension. To obtain a 
    cyclic solution, the
    magnitude of $H_5$ at the start of the big bang,
   $H_5({out})$,  must be
    slightly larger than the value at the end of the big crunch,
    $H_5({in})$. This is the case if more radiation
    is generated on the negative tension brane (see Appendix).
     }
     \end{figure}

To set the context for the later sections, we present a
brief tour through a single cycle, using the labels in 
Figure~1 as the mileposts.   Stage 1 represents the 
present epoch.  The current value of the 
Hubble parameter is $H_0=$~(15 billion yr)$^{-1}$.
We are presently at the time when  the scalar field is
acting as a form of quintessence in which its potential energy
has begun 
to dominate over matter and radiation.  Depending on the specific
details of the potential, the 
field $\phi$ may 
have already reached its maximal value (grey
circle), turned
back, and begun to evolve towards negative values.
If not, it will do so in the near future.
Because the slope of the potential is very small, 
$\phi$  rolls very slowly in the negative direction.
As long as the potential energy dominates, 
the Universe undergoes
 exceedingly slow cosmic acceleration (compared to inflationary
 expansion), roughly doubling in size every $H_0^{-1} =$~15
 billion years.  If  the acceleration
lasts trillions of years or more (an easy constraint to 
satisfy), 
the entropy and black hole
densities become negligibly small
 and the Universe
is nearly vacuous.
The Einstein equations become:
\begin{equation}
\label{eq1s}
H^2  =  \frac{8 \pi G}{3}
\left( \frac{1}{2} \dot{\phi}^2 +V(\phi)  \right)
\end{equation}
\begin{equation}
\frac{\ddot{a}}{a}  =  - \frac{8 \pi G}{3} \left(\dot{\phi}^2 -V(\phi)
 \right)
\label{eq2s}
\end{equation}
where $H$ is the Hubble parameter and $G$ is Newton's constant.
We will generally choose $8 \pi G=1$ except where otherwise noted.
Accelerated expansion  stops as 
$V(\phi)$ approaches zero and
 the scalar field kinetic energy becomes comparable to
the potential energy, Stage (2).  The  Universe continues to expand and 
the kinetic energy of scalar field continues to red shift
as the potential drops below zero. 
A nearly scale invariant spectrum of 
fluctuations on large length scales (beyond our current Hubble
horizon) begins to develop as the field rolls down the exponentially
decreasing part of the potential.  
The evolution 
 and perturbation
equations are the same as in the ekpyrotic model.\cite{kost1,ekperts}
Solving these equations, one finds that the decelerated expansion 
continues for a time $H_0^{-1}/c,$ which is
about one  billion years ($c$  is the parameter in 
$V(\phi)$, Eq.~(\ref{eq:examplep})).
At Stage (3), the potential becomes sufficiently negative
that the total scalar field energy density hits zero.
According to Eq.~(\ref{eq1s}), $H=0$  and 
the Universe is momentarily static.
From Eq.~(\ref{eq2s}), $\ddot{a} <0$, so that $a$
begins to contract. 
The Universe continues to satisfy the ekpyrotic conditions for
creating 
density perturbations.
Stage (4), about one second before the big crunch, is 
the
regime where fluctuations on the current Hubble horizon scale
are generated.  
As the field continues to roll towards $-\infty$,
the scale factor $a$ contracts and the 
kinetic energy of the scalar field grows.
That is, gravitational energy is converted to
scalar field (brane) kinetic energy during this part of the cycle.
Hence, the field races past the minimum of the potential at 
Stage (5) and
off to 
$-\infty$, with 
kinetic energy becoming increasingly dominant as the bounce
approaches.
The scalar field kinetic energy diverges as 
$a$ tends to zero.
At the bounce, Stage (6), matter and radiation are generated, the 
scalar field gets a kick and increases
speed as it reverses direction, and 
the Universe is expanding.
Through Stage (7),
the scalar kinetic energy density 
($\propto 1/a^6$) dominates over the radiation ($\propto 1/a^4$)
and the motion is almost exactly the 
time-reverse of the contraction phase between Stage (5) and the
big crunch.
 As the field rolls uphill, however, the 
 small kick given the scalar field and, subsequently,
 the radiation  become important, breaking the 
 time-reversal symmetry.
 The Universe becomes radiation dominated at  Stage (8), at say
$10^{-25}$~s after the big bang. 
The motion of $\phi$ is rapidly damped so that it converges towards 
its maximal value and then very slowly creeps downhill.
The damping continues during the matter dominated phase, which 
begins thousands of years later.
The Universe undergoes the standard big bang evolution 
for the next 15 billion  years, growing structure from the 
perturbations created when the scalar field was rolling downhill
at Stage (4).
Then, the  scalar field potential energy begins to dominate
and cosmic acceleration begins. 
Eventually, the scalar field rolls back across $\phi=0$.
The energy density falls to zero and cosmic
contraction begins. The scalar field rolls down the hill,
density perturbations are generated and $\phi$ runs off
to $-\infty$ for the next bounce. 
The evolution in terms of conventional variables is summarized
in Figure~\ref{avst}.  

\subsection{The View from the Visible Brane} \label{sec2c}

Thus far, we have described the evolution in terms of the usual
Einstein frame variables, $a$ and $\phi$.  However, as emphasized in the
next Section, these variables are singular at the transition from
big bang to big crunch, and they do not present an accurate 
picture of what an observer composed of matter confined to the brane
would actually see.  As $a$ approaches zero, the density of matter and
radiation
scale as $1/(a \beta(\phi))^3$  and $1/(a \beta(\phi))^4$,
respectively,
where $\beta(\phi)$ is a function of $\phi$ which scales as 
$1/a$ as $a$ tends to zero. Therefore the densities of
matter and radiation on the branes are  actually finite at $a=0$. 

This scaling of the density with $a \beta(\phi)$ rather than $a$ can be
understood rather simply. First, the spatial volume element on the
branes is that induced from five dimensions. When the brane
separation is small, one can use the usual formula for Kaluza-Klein
theory, 
\be \labeq{kk}
ds_5^2= e^{-\sqrt{2\over 3} \phi} ds_4^2 + e^{2\sqrt{2\over 3} \phi}
dy^2, 
\ee
where $ds_4^2$ is the four dimensional line element,
$y$ is the fifth spatial coordinate which runs from zero to $L$, 
and  $L$ is a parameter with the dimensions of length.
If we write the four dimensional line
element in conformal time coordinates, as $ds_4^2= a^2(-d\tau^2 +d\vec{x}^2)$,
then since from the Friedmann equation we have $(a'/a)^2 = {1\over 6} 
(\phi')^2$, we see  that $a$ is proportional to
$ e^{\phi/\sqrt{6}}$ in the big crunch. 
Hence  a three dimensional comoving volume element
$d^3x a^3 e^{-\sqrt{3\over 2}\phi}$  remains finite
as $a$ tends to zero. Thus the density of massive particles
tends to a constant. What about the density of radiation? First,
recall the usual
argument that the energy of a photon diverges at $a=0$. Consider
a set of comoving massive particles
in a spacetime with metric $a^2\eta_{\mu \nu}$ where
$\eta_{\mu \nu}$ is the Minkowski metric. The four velocities
of the particles obey $u^{\mu} u^{\nu} g_{\mu \nu}=-1$.
Hence, if they are comoving ($\vec{u}=\vec{0}$), then
we must have $u^0=a^{-1}$. Now a photon moving in
such a spacetime has a constant four-momentum,
$p^{\mu} = E(1,\vec{n})$, with $\vec{n}^2=1$. The energy
of the photon, as seen by the comoving particles, 
is $-u^\mu p^\nu g_{\mu \nu}=E/a$, which diverges
$a$ tends to zero. However, in the present
context, the metric to which the comoving particles
couple
is $e^{-\sqrt{2\over 3} \phi} a^2\eta_{\mu \nu}$.
Therefore,  we have $u^0=a^{-1}e^{\sqrt{1\over 6} \phi}$ and
the energy of the detected photons is finite as $a$ tends to zero. 
In other words, the effect of the scalar field  approaching
$-\infty$ is precisely such as to cancel the
gravitational blueshift.

The second crucial use of the higher dimensional metric
is in piecing together the global view of the spacetime. 
If one only had the Einstein frame scale factor $a$, it
would not be clear how to match to the next cycle, since
$a=0$ at the bounce. But the scale factor on 
a brane, $a \beta(\phi)$,
is nonzero at each bounce and may be so matched.
In fact, in the examples studied in this paper, the 
scale factors $a_0$ and $a_1$ (which are the brane scale factors
in the simplest  models) both undergo a net exponential
expansion within a cycle, and decrease for very brief periods
- either just before the brane collision (for $a_0$) or
just after it (for $a_1$). 
An observer on either brane would view the cosmology 
as one of almost uninterrupted expansion, with 
successive episodes of radiation-, matter-, and 
quintessence-domination ending in a sudden release of matter 
and radiation.

Both  matter and radiation are suddenly created by
the impact of the other brane.  The forewarning of this
catastrophic event would be that 
as $\beta(\phi)$ started to rapidly change, one would see
stronger and stronger violations of the equivalence principle
(a `fifth force'), and 
the masses and couplings of all particles would change.
In the case of M theory, the running of the string coupling
to zero would presumably destroy all bound states 
such as nucleons and send all particle masses to zero.

\section{Through the Bounce} \label{sec3}

To have repeating cycles,
the Universe must be able to pass smoothly from
a big crunch to a big bang.
Conventionally, the curvature and density 
singularity when the scale factor $a$ approaches 
zero  has been regarded
as an impassable obstacle to the understanding of
what came `before' the big bang. However, the brane world
setup sheds  new light on this problem. The key feature
is that the apparent singularity  in the
effective four-dimensional  description corresponds to 
a higher dimensional setup in which the four dimensional
metric is completely non-singular.
When the extra dimension (or outer brane separation) shrinks
to zero, there is no associated curvature singularity,
and the density of matter on the branes remains finite.
The most conservative assumption, based on the higher dimensional
picture,
is that the branes bounce from
(or, equivalently, pass through) each other and time continues
smoothly, with some conversion of brane kinetic energy to 
entropy. The separation of the two branes after the bounce 
corresponds to re-expansion in the four-dimensional effective theory.

How can this be reconciled with the singular 
four-dimensional description?
The point explained in 
Ref.~\ref{nonsing} is that the usual four-dimensional
variables, the scale factor $a$ and the scalar field $\phi$, are 
a singular choice at $a=0$. Each is poorly
behaved as the branes collide, but in the brane picture 
physical quantities depend on
combinations of the two variables that remain well-behaved.
These nonsingular 
variables may be treated as fundamental, and matching
rules derived to
parameterize the physics of inelastic
brane collisions. If the system can, as conjectured in Ref.~\ref{nonsing},
be properly embedded within string theory, the matching conditions 
will be derivable from fundamental physics.

\subsection{Non-singular variables}

The action for
a scalar field coupled to gravity and a set of fluids $\rho_i$
in a homogeneous, flat
Universe, with line element $ds^2=a^2(\tau)(-N^2 d \tau^2 +d\vec{x}^2)$ is
\be
{\cal S} = \int d^3x d \tau \left[
N^{-1}\left(-3a'^2+{1\over 2} a^2 \phi'^2\right)
-N \left(\left(a \beta\right)^4 \Sigma_i \rho_i + a^4 V(\phi))\right)\right].
\labeq{one}
\ee
We use  $\tau$ to represent
conformal time and  primes to  represent derivatives 
with respect to $\tau$.
$N$ is the lapse function. 
The background solution for the scalar field is denoted 
$\phi(\tau)$, and $V(\phi)$ is the scalar potential. 

The only
 unusual term in (\ref{eq:one}) is the coupling of the fluids
$\rho_i$, which we treat as perfect fluids coupled only through gravity.
The action for a perfect fluid coupled to gravity is just
$-\int d^4x \sqrt{-g} \rho$, where the density $\rho$ is regarded as
a function of the coordinates of the fluid particles and the
spacetime metric\cite{fock}. For a homogeneous
isotropic fluid, the equation of state $P(\rho)$ defines the 
functional dependence of $\rho$ on the scale factor $a$, via
energy-momentum conservation, $d {\rm ln} \rho/d {\rm ln} a =
-3(1+w)$, with $w=P/\rho$. For example, for radiation, $\rho \propto a^{-4}$
and for matter $\rho \propto a^{-3}$. 

We assume these fluids live on one of the branes, so that rather
than coupling to 
the Einstein-frame scale factor $a$, the particles they are composed of 
couple to 
a conformally related scale factor
$a \beta (\phi)$, being the scale factor on the appropriate brane.
For simplicity we have only written the action for fluids on one
of the branes, the action for fluids on the other brane being
a xerox copy but with the appropriate $\beta(\phi)$. 

  The function 
$\beta(\phi)$ may generally be different for the two branes,
and for different brane world setups. But as mentioned above there
is 
an important universality
at small separations corresponding to 
large negative $\phi$. In this limit, which is 
relevant to the bounce,
the bulk warp factor becomes irrelevant and 
one  obtains $\beta \sim e^{-\phi /\sqrt{6}}$, the
standard Kaluza-Klein result. This behavior ensures that
$ a \beta $ is finite at collision 
and so the matter and radiation densities
are, as well. 

The equations of motion for gravity, the matter and scalar field
$\phi$ are
straightforwardly derived by varying (\ref{eq:one}) with respect
to $a$, $N$ and $\phi$, after which $N$ may be set
equal to unity.  Expressed in terms of proper time
$t$,
The Einstein equations are
\begin{equation}
\label{eq1}
H^2  =  \frac{8 \pi G}{3}
\left( \frac{1}{2} \dot{\phi}^2 +V + \beta^4 \rho_R + 
\beta^4 \rho_M \right),
\end{equation}
\begin{equation}
\frac{\ddot{a}}{a}  =  - \frac{8 \pi G}{3} \left(\dot{\phi}^2 -V + 
\beta^4 \rho_R
+{1\over 2} \beta^4 \rho_M \right),
\label{eq2}
\end{equation}
where a dot is a proper time derivative.
As an example, we consider the case where there is
radiation ($\rho_R$)  and  matter ($\rho_M$) on the visible 
brane only, which could in principle be either the
positive or negative tension
brane. 
Then the above equations are  supplemented by the dynamical equation
for the evolution of $\phi$,
\begin{equation}
\label{eq2a}
\ddot{\phi}+3 H \dot{\phi} =  -V_{,\phi} - \beta_{,\phi} \beta^3 \rho_M
\end{equation}
and the continuity equation,
\begin{equation}
\hat{a} \frac{d \rho}{d \hat{a}} = a \frac{\partial \rho}{\partial a}
+ \frac{\beta}{\beta'} \frac{\partial \rho} {\partial \phi} = 
-3 (\rho + p)
\end{equation}
where $\hat{a} = a \beta(\phi)$ and
$p$ is the pressure of the fluid component with energy density 
$\rho$.
Note that only the matter density contributes to the $\phi$-equation,
because, if $\rho_R \propto 1/(a \beta)^4$, the radiation term is 
actually just a constant times $N$ in the action, contributing
to the Friedmann constraint but not the dynamical 
equations of motion. 

 If $\beta(\phi)$ is
  sufficiently flat near the current value of $\phi$,
    these couplings have modest effects in the late
    Universe, and the successes of the standard cosmology
     are recovered. For example the total variation in
     $\phi$  since nucleosynthesis is very modest. 
     In Planck units, this is of order $(t_r/t_N)^{1\over 2}$ 
     where $t_r$ is the time at the beginning of the 
     radiation dominated epoch 
     and  nucleosynthesis begins at 
     $t_N \sim $ 1 sec. It is utterly negligible for
     values of $t_r$ earlier than the electroweak era. 
     However, the coupling of matter to $\phi$
    produces other potentially measurable effects including
     a `fifth force' causing
violations of the equivalence principle.
 Current constraints can be satisfied if
   $M_{Pl} ({\rm ln} \beta)_{,\phi} <10^{-3}$.\cite{other,poly2,poly}

As the Universe contracts towards the big crunch, $a \rightarrow 0$,
the scalar field runs to 
$ -\infty$ and the scalar potential becomes 
negligible.  
The Universe becomes dominated 
by the scalar field kinetic energy density since it scales 
as $a^{-6}$ whereas matter and radiation
densities scale as $a^{-3}$ and $a^{-4}$ respectively (ignoring 
the $\beta$ factor). As 
scalar kinetic domination occurs, the scale factor 
$a$ begins to scale as  $(-t)^{1\over 3}$, and 
the background scalar field 
diverges logarithmically in time.
The energy density and
Ricci scalar diverge as $(-t)^{-2}$, so that $t=0$ is a 
`big crunch' singularity.


As explained in Ref.~\ref{nonsing}, in the simplest 
treatment of brane world models
there is only one scalar field modulus,
the `radion,' which runs off to minus infinity as 
the scale factor $a$ approaches zero.
The  singular variables, $a$ and $\phi$, 
can be replaced by the non-singular variables:
\be
a_0=2\,  a \,  {\rm cosh} ((\phi-\phi_{\infty})/\sqrt{6}) 
\qquad a_1=-2 \,  a \,  
{\rm sinh} ((\phi-\phi_{\infty})/\sqrt{6}).
\labeq{a0a1}
\ee
The kinetic terms in the action define the metric on moduli
space. In terms of the old variables one has the line
element $-3 da^2+{1\over 2} a^2 d\phi^2$, and $a=0$ is clearly
a singular point in these coordinates. However, in the new coordinates
in Eq.~(\ref{eq:a0a1}), the line element is 
${3\over 4}  (-da_0^2 +da_1^2)$, which is perfectly regular
even when the Einstein frame scale factor
$a={1\over 2} \sqrt{a_0^2-a_1^2}$ vanishes, on the `light-cone'
$a_0=a_1$. For branes in AdS, $a_0$ and $a_1$ are the scale factors
on the positive and negative tension branes\cite{kost1} so that
$\beta= 2 {\rm cosh} ((\phi-\phi_{\infty})/\sqrt{6}$ or
$-2{\rm sinh} ((\phi-\phi_{\infty})/\sqrt{6})$ respectively
for matter coupling to these branes. 

Notice that the constant field shift $\phi_{\infty}$ is arbitrary. Its
effect is a Lorentz boost on the $(a_0,a_1)$ moduli space. In 
the Kaluza-Klein picture (\ref{eq:kk}), a constant shift in $\phi$ 
can be removed by rescaling four dimensional spacetime coordinates
and re-defining the length scale $L$ of the extra dimensions. 
In the absence of matter which couples to $\phi$, or of a
potential $V(\phi)$,
this shift is unobservable, a reflection of the global symmetry
 $\phi \rightarrow \phi+$ constant of the 
4d effective theory. However, this symmetry
is broken by $V(\phi)$, and by matter couplings. In fact,
the scale
factor
$a_1$ must be positive in order for it to be 
interpretable  as a `brane scale factor', and this requires that 
$\phi< \phi_{\infty}$. 

We shall find it
convenient
to choose $\phi=0$ to be 
the zero of the potential $V(\phi)$, and then to 
choose $\phi_\infty$ so that 
$a_1$ never
vanishes for the solutions we are interested in. (In fact, 
since $a_1$ has a positive kinetic term~\cite{kost1}, 
a suitable coupling to moduli fields will always guarantee that
$a_1$ `bounces' away from zero~\cite{ekperts}.
In this paper, for simplicity
we ignore this complication by picking $\phi_\infty$ large enough
that no such `bounce' is necessary.)

Both $a_0$ and $a_1$ 
are `scale factors' since they transform like $a$ under rescaling space-time 
coordinates. However, unlike $a$ they tend to finite constants
as $a$ tends to zero, implying an alternative metrical
description which is not singular at the `big crunch'.
In the brane world models considered in Ref.~\ref{kost1},
$a_0$ and $a_1$ actually represent the 
scale factors of the positive and negative tension branes
respectively.
Since
there are no low energy configurations with $a_0<a_1$,
the `light cone' $a_0=a_1$ is actually a boundary of
moduli space and one requires a matching rule to 
determine what the trajectory of the system does at that point. 
A natural matching rule is to suppose that at low energies
and in the absence of potentials or matter, the branes 
simply pass through one another (or, equivalently, bounce)
with the intervening bulk
briefly disappearing and then reappearing after collision. 
This rule was detailed in Ref.~\ref{nonsing}, where simple models
satisfying the string theory background equations to all orders in
$\alpha'$ were given. In the Appendix we discuss 
the collision between boundary branes in terms of
energy and momentum conservation, and the Israel matching conditions.

Let us now comment on the character of the trajectory in the
($a_0$, $a_1$)-plane. The Friedmann constraint reads
\begin{equation}
{a}_0'^2 -{a}_1'^2 =  {4\over 3 } \left((a\beta)^4 \rho  + {1\over 16}
(a_0^2-a_1^2)^2 V(\phi_0)\right).
\label{eq:fc}
\end{equation}
If the energy density on the right hand side is positive,
the trajectory is time-like. If the right hand side
is zero (for example if the
potential vanishes as $\phi_0 \rightarrow -\infty$ and if there
is no matter or radiation), then the trajectory is light-like.
If the right hand side is negative, the trajectory is space-like.

\begin{figure} \label{fig3}
\begin{center}
 \epsfxsize=5.0 in \centerline{\epsfbox{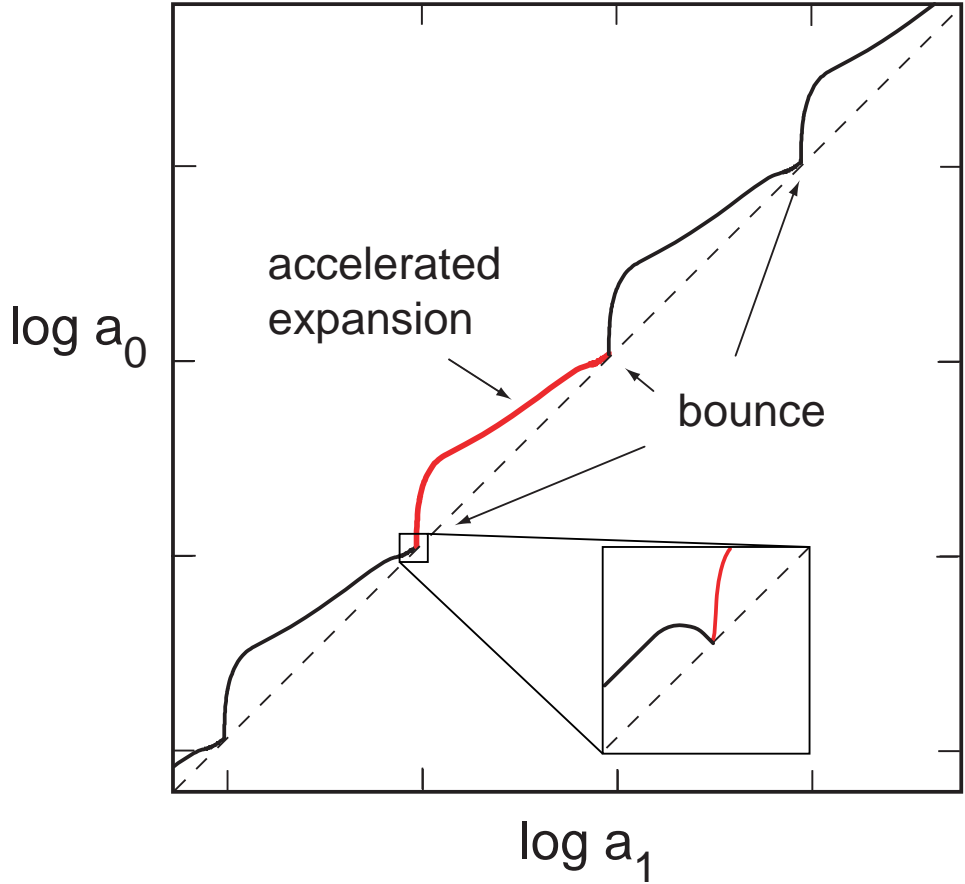}}
   \end{center}
      \caption{
 Schematic plot of the $a_0$-$a_1$ plane showing a sequence of
   cycles of expansion and contraction (indicated by
  tick marks).  The dashed line represents
 the ``light-cone" $a_0 =a_1$ corresponding to a bounce ($a=0$).
  Each cycle includes a moduli kinetic energy, radiation, matter and
quintessence dominated phase and lasts an exponentially large number
   of e-folds. The insert shows the trajectory near the big crunch
     and bounce. The potential energy $V(\phi)$ assumed takes the
      form shown in Fig.~1.
      }
      \end{figure}

The trajectory for the cyclic solution in the $a_0-a_1$ plane
is shown in Figure~\ref{fig3}.  
The insert shows a blow-up of the behavior at the bounce in which 
the trajectory is light-like at contraction to the big
crunch (the Universe 
is empty) and time-like on expansion from the big bang 
(radiation is produced 
at the bounce).
In these coordinates, the scale factor increases exponentially 
over each cycle, but the next cycle is simply a rescaled 
version of the cycle before. A local observer  measures
physical quantities such as 
the Hubble constant or the deceleration parameter,
which entail ratios of the scale factor and its derivatives 
in which the normalization of the scale factor cancels out.
Hence, to local observers, 
each cycle appears to be identical to the one before.

\subsection{From Big Crunch to Big Bang}

In this section we solve the equations 
of motion immediately before and after the bounce,
and discuss how the incoming and outgoing states are
connected. The nonsingular `brane' scale factors $a_0$
and $a_1$ provide the natural setting for this discussion,
since neither vanishes at the bounce. As emphasized
above, the Einstein frame scale factor $a$, and the
scalar field $\phi$ are singular coordinates on field
space at the bounce. Nevertheless, since our intuition 
is much better 
in the Einstein frame, we shall also give formulae for
$a$ and $\phi$ near the bounce.  In subsequent 
sections,  we shall frame the discussion almost 
entirely in terms of Einstein frame variables, for
the most part using 
the nonsingular
variables $a_0$  and $a_1$ solely as a `bridge' connecting the
incoming big crunch to the outgoing big bang. 

Before the bounce there is little radiation present since
it has been 
 exponentially diluted in the preceding quintessence-dominated 
accelerating phase. Furthermore, the potential $V(\phi)$ becomes negligible
as 
 $\phi$ runs off to minus
infinity.
The Friedmann constraint reads $(a'/a)^2 = {1\over 6} \phi'^2$,
and the scalar field equation, $(a^2 \phi')'=0$, where primes denote
conformal time derivatives. The 
general solution is 
\ba
\phi&=& \sqrt{3\over 2} {\ln \left(A H_5({in}) \tau \right),  \qquad   
a=A e^{\phi/\sqrt{6}}} = A \sqrt{A H_5({in}) \tau },\cr
a_0&=&A\left(\lambda+\lambda^{-1}A H_5({in}) \tau \right),
 \qquad a_1=A\left(\lambda-\lambda^{-1}A H_5({in}) \tau \right), 
\label{eq:before}
\ea
where $\lambda\equiv e^{\phi_{\infty}/\sqrt{6}}$. We choose 
$\tau =0$ to be the time when $a$ vanishes so that
$\tau<0$ before collision.
$A$ is an integration constant
which could be set to unity by rescaling space-time coordinates
but it is convenient not to do so. The Hubble constants as defined in terms
of the brane scale factors are $a_0'/a_0^2$ and $a_1'/a_1^2$ which
at $\tau=0$ take the values 
$+\lambda^{-3}H_5({in})$ and $-\lambda^{-3}H_5({in})$ respectively.

Re-expressing the scalar field as a function of proper time $t = \int a d\tau$,
we obtain
\be
\phi=\sqrt{2\over 3} {\rm ln} \left({3\over 2} H_5({in})t\right).
\labeq{ksol}
\ee
The integration constant  $H_5({in})<0$
has a 
natural physical interpretation as a measure of the 
contraction rate of the extra dimension (See equation (\ref{eq:kk})):
\be \label{defh5}
H_5 \equiv \frac{d L_5}{L dt_5} \equiv
\frac{d \, ( e^{\sqrt{2\over 3} \phi})}{d t_5} = \sqrt{\frac{2}{3}} \dot{\phi}
 e^{\sqrt{3\over 2} \phi},
\ee
where $L_5 \equiv L e^{\sqrt{2\over 3} \phi}$ is the proper
length of the extra dimension,    
$L$ is a parameter with dimensions of length, and 
$t_5$ is the proper time in the five-dimensional metric,
\be 
d t_5 \equiv a  e^{-\sqrt{1\over 6} \phi} d \tau = e^{-\sqrt{1\over 6} \phi} dt,
\ee
with $t$ being FRW proper time. Notice that a shift $\phi_\infty$ can
always be compensated for by a rescaling of $L$. 
As the extra dimension shrinks to zero, 
$H_5$ tends to
a  constant, $H_5({in})$. 

If the extra dimension shrinks adiabatically and backreaction
from particle production can be ignored, then the matching rule
conjectured in Ref.~\ref{nonsing} states that 
$H_5$ after the bounce should be given by
$H_5(out)=-H_5(in)$.
However, if radiation is produced, $H_5(out)$ takes 
a different value. If one is given the densities of radiation
produced on both branes, then $H_5(out)$ may be inferred 
from energy and momentum conservation,
and the Israel matching conditions, as we show in the Appendix.

Immediately after the bounce, scalar kinetic energy
dominates and 
$H_5$ remains nearly constant,
as shown in Figure~\ref{avst}.
The kinetic energy of the scalar field scales as
$a^{-6}$ and  radiation scales  as $a^{-4}$, so the former
dominates at small $a$. 
It is convenient to re-scale $a$
so that it is unity at scalar kinetic energy-radiation equality, $t_r$,
and denote the corresponding Hubble constant $H_r$.
The Friedmann constraint in Eq.~(\ref{eq:fc}) then reads
\be
(a')^2 = {1\over 2} H_r^2 (1+ a^{-2}),
\labeq{fcon}
\ee
and the solution is
\ba
\phi&=&\sqrt{3\over 2} {\rm ln}\left({ 2^{5\over 3}  \tau
H_5^{2\over 3} ({out}) H_r^{1\over 3}
\over (H_r \tau +2^{3\over 2} )}\right),\qquad 
a=\sqrt{{1\over 2} H_r^2 \tau^2  +\sqrt{2} H_r \tau}.
\labeq{after}
\ea
The brane scale factors are
\ba
a_0&\equiv& a(\lambda^{-1}e^{\phi/\sqrt{6}} +\lambda e^{-\phi/\sqrt{6}}) =
A \left(\lambda(1+{H_r \tau
\over 2^{3\over 2}}) + \lambda^{-1} 2^{1\over 6} H_r^{1\over 3} H_5^{2\over 3} ({out})
 \tau \right),\cr
a_1&\equiv&
a(-\lambda^{-1}e^{\phi/\sqrt{6}} +\lambda e^{-\phi/\sqrt{6}})\cr
&=&
A \left(\lambda(1+{H_r \tau
\over 2^{3\over 2}}) -  2^{1\over 6} \lambda^{-1} H_r^{1\over 3} H_5^{2\over 3} ({out})
 \tau \right).
\labeq{a0a1rad}
\ea
Here the constant $A=2^{1\over 6} \left(H_r/ H_5({out}) \right)^{1\over 3}$
has been defined so that we match $a_0$ and $a_1$ to
the incoming solution given in 
(\ref{eq:before}). As for the incoming solution,
 we can compute the
Hubble constants on the two branes after collision. They
are $\pm \lambda^{-3} H_5(out) + 2^{-{5\over 3}}  \lambda^{-1} H_r^{2\over 3} 
H_5^{1\over 3} $ on
the positive and negative tension branes respectively.

For $H_r <2^{5\over 2} \lambda^{-3} H_5$, the case of relatively
little radiation production, immediately after collision 
$a_0$ is expanding but $a_1$ is contracting. Whereas for 
$H_r >  2^{5\over 2} \lambda^{-3} H_5$, both brane scale factors
expand after collision.
We shall concentrate on the 
former case in this paper, in which we are near the adiabatic limit. 
If no scalar potential $V(\phi)$ were present,
the scalar field would continue to obey the solution
(\ref{eq:after}),  converging to 
\be
\phi_C=\sqrt{2\over 3} {\rm ln}\left(2^{5\over 2}  
{H_5({out})\over H_r}\right).
\labeq{aftercon}
\ee
This value is actually larger than $\phi_\infty$ for
$H_r <H_5\lambda^{-3} 2^{5\over 2}$, the case of weak
production of radiation. However, 
the presence of the potential $V(\phi)$ alters the
expression 
(\ref{eq:aftercon}) for the final resting value of
the scalar field.
As $\phi$ crosses the potential well traveling in the 
positive direction,  
$H_5$ is reduced to a renormalized value $\hat{H}_5({out})
<H_5({out})$, so that the final resting 
value of the scalar field can be smaller than $\phi_\infty$.
If this is the case, then $a_1$ never crosses zero,
instead reversing to expansion shortly after radiation 
dominance. (In the 
calculations of Ref~\ref{ekperts}, where we assumed
the potential {\it vanished} after collision, this
effect did not occur. Instead, we invoked a coupling of
$a_1$ to a modulus field which caused it to bounce off
$a_1=0$.).

If radiation dominance occurs 
well after $\phi$ has crossed the potential well, 
Eq.~(\ref{eq:aftercon}) provides a reasonable estimate
for the final resting value, if we use the 
corrected value $\hat{H}_5({out})$. The dependence of
(\ref{eq:aftercon}) is simply understood: while 
the Universe is kinetic energy dominated, 
$a$ grows as 
$t^{1\over3}$ and $\phi$ increases logarithmically
with time.
However, when the Universe becomes radiation-dominated
and $a\propto t^{1\over 2}$, 
Hubble damping increases and $\phi$ converges to the finite
limit above.

\section{Across the Well} \label{sec4}

Using the potential described in Section~\ref{sec2a} and,
specifically, the example
 in Eq.~(\ref{eq:examplep}),
this Section considers the motion of $\phi$ 
back and forth across the potential well. We will  show
that evolution converges to a stable attractor solution.
Our main purpose, though, 
is to explore the asymmetry in the behavior before and after
the bounce
that is an essential component of the cyclic solution.

Over most of this region, $V$ may be accurately 
approximated by 
$ -V_0 \, e^{-c \phi}$. For this pure exponential potential,
there is a simple scaling solution\cite{ekperts}
\be
a(t)= |t|^p, 
\qquad V=-V_0e^{-c \phi} = -{p(1-3p)\over t^2},
\qquad 
p={2\over c^2},
\labeq{back}
\ee
which is an expanding or contracting 
Universe solution according to whether  $t$ is positive or negative.
(We choose $t=0$ to be the bounce.) From the expression for $V$,
we see that $\phi$ varies 
logarithmically with $|t|$.

At the end of the expanding phase
of the cyclic scenario, there is a period of 
accelerated expansion which  makes the Universe empty, 
homogeneous and flat, followed by $\phi$  rolling down the 
potential $V(\phi)$ into the well. After $\phi$ has 
rolled sufficiently and the scale factor has begun to
contract (past Stage~(3) in Figure~\ref{potent}),
the Universe accurately follows the above scaling solution
down the well until
$\phi$ encounters the potential minimum (Stage (5) in  
Figure~\ref{potent}).

Let us consider the behavior of $\phi$ under small shifts
in the contracting phase.
In the background scalar field equation and the
Friedmann equation, we set 
$\phi=\phi_B+\delta \phi$ and $H=H_B+\delta H$, where $\phi_B$ and
$H_B$ are the background quantities given from (\ref{eq:back}).
To linear order in 
$\delta \phi$, one obtains 
\be
\delta \ddot{\phi} +{1+3p\over t} \delta \dot{\phi} - {1-3p\over t^2} \delta \phi =0,
\labeq{pfq}
\ee
with two linearly independent solutions, 
$\delta \phi \sim t^{-1}$ and $t^{1-3p}$, where $p \ll 1$. 
In the contracting phase, the former solution grows 
as $t$ tends to zero.  However, this solution is simply an infinitesimal
shift in the time to the big crunch: $\delta \phi \propto \dot{\phi}$.
Such a shift provides a solution to the 
Einstein-scalar equations because they are time translation 
invariant, but it is 
physically irrelevant since it can be removed by
a redefinition of time.
The second solution is a physical perturbation mode and it
decays 
as $t$ tends to zero. Hence, we find that  the background 
solution is  an attractor in the contracting phase. 

We next consider the incoming and outgoing collision velocity,
which we have parameterized as $H_5({in})$ and $H_5({out})$ in the 
previous Section.
Within the scaling  solution (\ref{eq:back}), we can calculate
the value of incoming velocity by treating  the prefactor of
the potential
$F(\phi)$ in  Eq.~(\ref{eq:examplep})
as a Heaviside function which is unity for $\phi > \phi_{min}$
and zero for $\phi < \phi_{min}$, where $\phi_{min}$ is the
value of $\phi$ at the minimum of the potential.
We compute the velocity of the field as it approaches $\phi_{min}$ and
use energy conservation at the jump in $V$ to infer the velocity
after $\phi_{min}$ is 
crossed. 
In the scaling solution, the total energy as $\phi$ approaches 
$\phi_{min}$ from the right is
$\frac{1}{2} \dot{\phi}^2 +V  = 3 p^2/t^2$, and this must equal
the total energy $\frac{1}{2} \dot{\phi}^2$ evaluated for 
$\phi$ just to the left of 
 $\phi_{min}$. Hence, we find that $\dot{\phi} = \sqrt{6}p/t =
 \sqrt{6 p V_{min}/(1-3p)}$ at the minimum 
 and, according to Eq.~(\ref{defh5}), 
\be
H_5({in})\approx -{\sqrt{8}\over c} {|V_{min}|^{1\over 2} 
e^{\sqrt{3\over 2}
\phi_{min}}\over \sqrt{1-6 c^{-2}} }.
\labeq{h5in}
\ee
At the bounce, this solution is matched to an expanding solution with
\be
H_5({out}) = -(1+\chi) H_5({in})>0,
\labeq{h5mat}
\ee
where $\chi$ is a small parameter which arises because of the
inelasticity of the collision.

In order to obtain cyclic behavior, 
we shall need  $\chi$ to be positive or, equivalently,
the outgoing velocity to exceed the
incoming velocity. 
There are at least two effects that can cause $\chi$ to be positive.
 First, as we discuss in the
Appendix,   $\chi$ is  generically 
 positive if more radiation is
generated on the negative tension brane 
than on the positive tension brane at collision.
Secondly, $\chi$ can get a positive contribution from 
the coupling of $\beta(\phi)$ to the matter created on the branes
by the collision;
see Eq.~(\ref{eq2a}).
Both effects are equally good for our purposes.
For the present discussion,  we shall simply assume a small
positive $\chi$ is given, and follow the evolution forwards
in time.

Since $\chi$ is small, 
the outgoing solution is very nearly the
time reverse of the incoming solution
as $\phi$ starts
back across the potential well
 after the bounce:
the  scaling solution given in (\ref{eq:back}),
but with $t$ positive. As time proceeds however, the 
contribution of $\chi$ becomes increasingly significant.
In the time-reversed
scaling solution, 
$H_5$ tends to zero.
For $\chi>0$  $H_5$ remains positive and $\phi$ overshoots 
the potential well. 
$V_0$ is exponentially smaller than the kinetic energy 
density at the bounce, so even a tiny fraction $\chi$ suffices
to reach the plateau after crossing the potential well.

We can analyze this overshoot by treating $\chi$ 
as a perturbation and using the solution in Eq.~(\ref{eq:pfq}) 
discussed above, 
$\delta \phi \sim t^{-1}$ and $t^{1-3p}$. 
The latter is a decaying mode in the contracting phase before
the bounce but it grows
in the expanding phase.
One can straightforwardly compute the perturbation
in $\delta H_5$ in this growing mode by matching at $\phi_{min}$
as before. One finds $\delta H_5=12 \chi H_5^B/c^2$ where 
$H_5^B$ is the background value, at the minimum. Beyond this
point, 
$\delta H_5$ grows as  $t^{\sqrt{6}/c} 
\propto e^{\sqrt{3\over 2} \phi}$, for large $c$, whereas
in the background scaling solution 
$H_5$ 
decays with $\phi$ as $e^{(\sqrt{3\over 2}-c/2)\phi}$.
When the perturbation is of order the background value, 
the trajectory departs from the scaling solution and the 
potential becomes irrelevant. The departure occurs when the scalar
field has attained the value
\be
\phi_{Dep} = \phi_{min}
+ {2\over c} {\rm ln} {c^2\over12 \chi},
\qquad
|V| \lsim  \left({12 \chi\over c^2}\right)^2 |V_{min}|.
\labeq{pertbig}
\ee
As $\phi$ passes beyond  $\phi_{Dep}$ the kinetic energy 
overwhelms the negative potential and 
the field passes onto the plateau $V_0$ with 
$H_5$ nearly constant (see Figure~\ref{avst}), and equal to
\be
\hat{H}_5 ({out}) \approx \chi 
\left({c^2\over 12 \chi}\right)^{\sqrt{6}\over c}
H_5({in}), 
\labeq{apph5}
\ee
until the radiation, matter and vacuum energy become
significant and $H_5$ is then damped away to zero.

Before moving on to  discuss these late stages, it
is instructive to compare how rapidly $\phi$ travels
over its range before and after the bounce. 
The time spent to the left of the potential well ($\phi < \phi_{min}$)
is essentially identical in the
incoming and outgoing stages for $\chi <<1$, namely
\be
|t_{min}|\approx {c \over 3 \sqrt{2 |V_{min}|}}.
\labeq{timefmin}
\ee

For the outgoing solution, when $\phi$ has left the scaling solution
but before radiation domination, the definition 
Eq.~(\ref{defh5}) may be integrated to give 
the time since the big bang at each value of $\phi$,
\be
t(\phi) = \int {d\phi \over \dot{\phi}}
= \sqrt{2\over 3} \int d \phi {e^{\sqrt{3\over 2} \phi} \over H_5(\phi)}
\approx {2\over 3} { e^{\sqrt{3\over 2} \phi} \over \hat{H}_5({out})}.
\labeq{timeo}
\ee
The time
in Eq.~(\ref{eq:timeo}) is a {\it microphysical}
scale.
The corresponding formula for 
the time before the big crunch is very different.
In the scaling solution (\ref{eq:back}) one has for large $c$
\be
t(\phi) = - \sqrt{2\over |V_{min}| } {e^{c(\phi-\phi_{min}) /2} \over c}
= -{6e^{c(\phi-\phi_{min})/2} \over c^2} |t_{min}|.
\labeq{timein}
\ee
The large exponential factor
makes the time to the big crunch far longer than the time from
the big bang, for each value of $\phi$. This effect is due
to the  increase in
$H_5$ after the bounce,  
which, in turn, is due to the positive value of $\chi$.

\section{The Radiation, Matter and Quintessence Epochs} \label{sec5}

As the scalar field passes beyond the potential
well, it runs onto the positive plateau $V_0$. 
As mentioned in the last section, the value of
$H_5({out})$ is nearly canceled in the passage 
across the potential well, and is reduced to 
$\hat{H}_5$ given in Eq.~(\ref{eq:apph5}). Once radiation domination begins,
the field quickly  converges to 
the large $t$ (Hubble-damped) limit of Eq.~(\ref{eq:after}), namely
\be
\phi_C=
\sqrt{2\over 3} {\rm ln} \left(2^{5\over 2}  \hat{H}_5({out})/H_r) \right),
\labeq{ksolc}
\ee
where $H_r$ is the Hubble radius at kinetic-radiation equality. 
The dependence is obvious: the asymptotic value of $\phi$ 
depends on the ratio of $\hat{H}_5({out})$  to $H_r$.
Increasing 
$\hat{H}_5({out})$ pushes $\phi$ further, likewise
lowering $H_r$ delays 
radiation domination allowing the 
logarithmic growth of $\phi$ in the kinetic energy dominated phase
to continue for
longer. 

As the kinetic energy red shifts away,
the gently sloping potential gradually becomes important,
in acting to slow and 
ultimately reverse $\phi$'s motion. 
The solution of the scalar field equation is, after expanding 
Eq.~(\ref{eq:after}) for large $\tau$,
converting to proper time 
$t=\int a(\tau) d \tau$ and matching,
\be
\dot{\phi}\approx {\sqrt{3} H_r \over a^3(t)} -a^{-3} \int_0^t dt a^3
V_{,\phi},
\labeq{solf}
\ee
where as above we define $a(t)$ to be unity at
kinetic-radiation equal density. 
During the radiation and matter eras, the first term scales as
$t^{-{3\over 2}}$ and $t^{-2}$ respectively.
For a slowly varying field,
$V_{,\phi}$ is nearly constant,
and the potential gradient
term in Eq.~(\ref{eq:solf}) scales linearly with $t$, so 
it eventually dominates. 

When does $\phi$ turn around? We give a rough discussion
here, ignoring factors of order unity. First, 
we use the fact that $V_0 \sim t_0^{-2}$ 
where $t_0$ is the present age of
the Universe, and roughly we have $t_0 \sim 10^{5} t_m$, where
$t_m$ is the time of matter domination. As we shall see,
$\phi$ may reach its maximal value $\phi_{max}$
and 
turn around during the radiation, matter or quintessence 
dominated epoch.  All three
possibilities are  acceptable phenomenologically, although the 
case where turnaround occurs in the radiation epoch appears 
more likely.
For example, 
$\phi_{max}$ is reached in the radiation era, if, from 
Eq.~(\ref{eq:solf}), 
\be
{t_{max} \over t_m } 
\approx 10^4 \left({t_r\over t_m}\right)^{1\over 5}
\left({V \over V_{,\phi}}(\phi_C)\right)^{2\over 5} <1.
\labeq{raddom} 
\ee
If the 
Universe becomes radiation dominated at the GUT scale,
$t_r \sim 10^{-25}$ seconds. Then, only if we fine-tune
such that $V/V' > 10^{10}$
does $t_{max}$ exceed $t_m$.  
This corresponds to the case where  we have  $10^{10}$ 
 e-foldings  or more of cosmic acceleration
at late times as $\phi$ rolls back, far 
more than required for the cyclic solution.
The bound changes somewhat if
 the Universe becomes radiation dominated as late as 
nucleosynthesis
($t_r \sim 1$ second).  In that
case,  even if $V/V_{,\phi}((\phi_C)$ is not much 
greater than unity, the scalar field turns around in the matter era
or later. For turnaround in the matter era, we require 
\be
3 \times 10^{-4} \lsim  \left({t_r\over t_m}\right)^{1\over 6}
\left({V \over V_{,\phi}}(\phi_C)\right)^{1\over 3}\lsim 30.
\labeq{matdom} 
\ee
Finally, 
if the field runs to very large 
$\phi_C$, so that $V_{,\phi}/V (\phi_C)\approx ce^{-c\phi_C}$
is exponentially small,
then $\phi$ only turns around in the quintessence-dominated era. 
For the example considered here, the natural range 
of parameters corresponds to turnaround occurring during the 
radiation-dominated epoch. Hence, by the present epoch, the field is
rolling monotonically in the negative direction and
slowly gaining in speed.
Consequently, the ratio of the pressure to the energy density is
increasing from its value at turnaround, $w=-1$, towards zero.
Depending on the details of the scalar potential $V(\phi)$, it
is conceivable that the
increasing value of $w$ could ultimately be observationally detectable.

Once the field has turned around and started to roll
back towards the potential well, the second 
term in  Eq.~(\ref{eq:solf}) dominates. 
For our scenario to be
viable, we require there to be a substantial epoch of 
vacuum energy domination (inflation) before the next big crunch.
The  number of e-foldings $N_e$ of inflation 
is given by the usual slow-roll formula,
\be
N_e = \int d\phi {V\over V_{,\phi}} \approx {e^{c \phi_C}\over c^2},
\labeq{slowroll}
\ee
for our model potential.
For example, if we
demand that the 
number of baryons per 
Hubble radius be diluted to 
below unity before the next contraction, which
is certainly over-kill in guaranteeing  that
the cyclic solution is an attractor, we set
$e^{3 N_e} \gsim 10^{80}$, or $N_e\gsim 60$.
This is easily fulfilled if $\phi_C$ is of order unity
in Planck units. 

From the formulae given above
we can also calculate the maximal value 
$\phi_C$ in the cyclic solution: for large $c$ and 
for $t_r>> \chi^{-1} t_{min}$, it is 
\be
\phi_C-\phi_{min}\approx 
\sqrt{2\over 3} {\rm ln} \left(\chi {t_r\over t_{min}}\right),
\labeq{cycrst}
\ee
where we used $H_r^{-1}  \sim t_r$, the beginning of the  
radiation-dominated epoch.
From Eq.~(\ref{eq:cycrst}) we obtain
\be
{t_r\over t_{min}} \sim  {1\over \chi} 
\left({c^2 N_e |V_{min}| \over V_0} \right)^{\sqrt{3\over 2 c^2}}. 
\labeq{trtm}
\ee
This equation provides a lower bound on  $t_r$.
The extreme case is to take 
$|V_{min}| \sim 1$. Then using $V_0 \sim
10^{-120}$, 
$c \sim 10$, $N_e \sim 60$, we find $t_r \sim 10^{-25}$ seconds.
In this case the maximum temperature of the Universe 
is $\sim 10^{10}$ GeV. This is not very different to what 
one finds in simple inflationary models.


As $\phi$ rolls down the hill, one
can check that $\phi$ leaves the slow-roll
regime when $e^{-c\phi}$ exceeds $3/c^2$. At this 
point the 
constant term $V_0$ in the potential becomes irrelevant 
and one can use the 
scaling solution for $\phi$, all the way to the potential minimum.
This is also the point at which density perturbations
start to be generated via the ekpyrotic mechanism,
while the Einstein frame scale factor $a$ is still expanding.
The Universe continues
to expand slowly, but with a slowly decreasing Hubble constant,
and finally enters contraction when 
when the density in the scalar
field reaches zero, at a negative value of the potential energy.
The ensuing contracting phase is accurately described by
the scaling solution (\ref{eq:back}), in which 
$a \sim (-t)^p$ and
$\dot{\phi} = \sqrt{2p} /t$, with
$t<0$, and $t=0$ being the time of the next big crunch. 
From the formulae 
(\ref{eq:a0a1}) 
one finds 
$\dot{a_1} = (p a_1  -\sqrt{p/3} a_0)/t$, which is greater than 
zero for $p<{1\over 3}$, since 
$a_0$ is greater than $a_1$. Thus even when $a$ is undergoing slow
contraction, in the scaling era, the effect of the motion of
$\phi$ is enough to make 
$a_1$ expand throughout this phase. Matter residing on this
brane would see continuous expansion all the way to the big crunch.
The same argument shows that
$a_1$ actually undergoes a small amount of 
contraction in the very much shorter scaling
epoch of the expanding phase.

\section{Generation of Density Perturbations}
\label{sec6}

In the cyclic scenario, the period of 
exponential expansion occurring late in each cycle plays a key
role in diluting the densities of matter, radiation and 
black holes to negligible levels, suppressing
long wavelength perturbations and establishing a
`clean slate', namely a flat vacuous Universe in which
all fields are in their quantum mechanical ground state.
As the scalar field rolls down the potential in 
Eq.~(\ref{eq:examplep}), entering the scaling solution
in Eq.~(\ref{eq:back}), the ekpyrotic mechanism for the
generation of fluctuations derived in Refs.~1 and
\ref{ekperts} sets in and a scale invariant spectrum
of adiabatic perturbations is thereby developed.
Quantum fluctuations of the usual inflationary sort
are also developed in the slow-roll quintessence
epoch, but these are: (a) negligible in
amplitude because $V_0$ is tiny;
and, (b) only excited on scales of order $t_0$ and above
in the contracting phase. These scales are
shrunk only as $(-t)^p$ in the contracting, scaling solution,
but then expanded as $t^{1\over 3}$, $t^{1\over 2}$ and
$t^{2\over 3}$ in the kinetic dominated, radiation and
matter eras in the big bang phase, which also lasts for
a time of order $t_0$. Therefore, the modes amplified 
during inflation are exponentially larger in wavelength than the 
Hubble radius scale in the next cycle by time of quintessence
domination, which is the present epoch.

Let us concentrate on the fluctuations produced
via the ekpyrotic mechanism.\cite{ekperts}
Expanding the inhomogeneous  fluctuations in the scalar
field $\delta \phi(t,\vec{x})= \Sigma_{\vec k} \delta\phi_{\vec{k}}(\tau)
e^{i \vec{k}\cdot \vec{x}}$, we remove the damping term by setting 
$\delta \phi_{\vec{k}}= a^{-1} \chi_{\vec{k}}$, to 
obtain
\be
{\chi}''_{\vec{k}}= -k^2 {\chi}_{\vec{k}} + \left({a''\over a}
-V_{,\phi\phi} a^2\right) {\chi}_{\vec{k}}\equiv
-\left(k^2- k_F^2\right) {\chi}_{\vec{k}},
\labeq{perts}
\ee
where primes denote conformal time derivatives and
we have defined $k_F$, the comoving 
`freeze-out' wavenumber. Modes with $k>k_F$ oscillate
with fixed amplitude, whereas those with $k<k_F$ are 
amplified. In the regime of interest $k_F$
grows monotonically so that shorter and shorter wavelengths
progressively freeze out as the big crunch is approached.
The physical scale at which modes freeze out is given by
\be
\lambda_{F} = \left[{a''\over a^3} -V_{,\phi\phi}\right]^{-{1\over 2}} =
 \left[{2\over 3} V -{1\over 6} \dot{\phi}^2 -V_{,\phi\phi}\right]^{-{1\over2}},
\labeq{free}
\ee
As usual we adopt units where $8 \pi G=1$, and denote
proper time derivative with dot. 
In the era of quintessence domination
when $V$ dominates over $V_{,\phi\phi}$, the freeze-out scale
$\lambda_{F}$
is nearly constant, and comoving wavelengths are exponentially
stretched beyond it.
As $V_{,\phi\phi}$ begins to dominate however, 
Hubble damping becomes irrelevant, and the system approaches the 
scaling solution given in Eq.~(\ref{eq:back}), in which
$V_{,\phi \phi} \approx - 2/t^2$, where $t$ is the proper time to
the big crunch. The freeze-out scale drops linearly with time to zero,
as the scale factor is falling, like $(-t)^{1\over 3}$.
Therefore progressively shorter and shorter wavelength modes
are frozen out and amplified, with waves of physical wavelength
$t_{F}$ being frozen out at a time $t_{F}$. 

An exponentially large band of comoving 
wavelengths is amplified and frozen in as $\phi$ rolls 
from $\phi=0$ down towards $\phi_{min}$. 
Modes with all physical wavelengths from the microphysical
scale $t_{min}$, which could be not much larger than the
Planck length, to the macroscopic scale $t_0/c$ which is
of order a tenth the present Hubble radius, aquire 
scale invariant perturbations. 
Once the perturbations
are generated, their wavelength scales as
$(t_{min}/t_{F})^p$ in the collapsing phase. 
Then as $\phi$ crosses the potential well and races off to
minus infinity, the Einstein frame physical wavelength 
goes to zero. But this is not the relevant quantity to
track, since we match the variables $a_0$ and $a_1$ and therefore
should match the physical wavelengths as measured by 
these scale factors.  In the kinetic dominated phase,
$a_0$ and $a_1$ are nearly constant, so in effect the
physical wavelength of the modes are matched 
when $\phi$ crosses $\phi_{min}$, in the contracting and
expanding phases. Furthermore, the contracting and expanding
solutions are  nearly time-reverses of one another,
until the time $t_{Dep}$ computed above when the expanding
solution deviates from scaling.  Therefore one is
effectively matching at $t_{Dep}$, from which one sees that 
the time $t_F$ at which perturbations on the
current Hubble radius scale $t_0$ were generated, is given by
\be
|t_F| \left({|t_{Dep}|\over |t_F|}\right)^p  \approx t_0 
\left({t_m \over t_0}\right)^{2\over 3} 
\left({t_r \over t_m}\right)^{1\over 2} 
\left({t_{Dep} \over t_r}\right)^{1\over 3}, 
\labeq{lamb}
\ee
where the bracketed factors are: (a) the contraction of
the scale factor in the scaling solution,
between the time $t_F$ at which the perturbations were
generated and the time $t_{Dep}$ at which the 
expanding solution departs from scaling; (b) the
scaling back of the present comoving Hubble radius scale 
to the time of matter-domination $t_m$; 
(c) the scaling back to the time of radiation-domination $t_r$;
and (d) the scaling back to the time $t_{Dep}$ 
using $H_5 \sim$~constant, corresponding to kinetic domination
in the expanding solution. 

From 
Eq.~(\ref{eq:timein}), 
it follows that 
perturbations on the scale of the 
 present Hubble radius were generated at a field value
\be
\phi_{F}\approx \phi_{min} +{2\over c} {\rm ln} \left(
{c^2 \over 6} \left({c^2\over 12 \chi}\right)^{1\over 3} 
{t_0^{1\over 2} t_r^{1\over 6} \over t_{min}^{2\over  3}}\right)
\labeq{curent}
\ee
Comparing with Eq.~(\ref{eq:cycrst}) for
the resting value of the field $\phi_C$, and the 
expression 
\be 
\phi_{min}\approx  - {2\over c} {\rm ln} \left(c t_0 /t_{min}\right),
\labeq{minf}
\ee
for the field value at the potential minimum, which follows
from Eq.~(\ref{eq:timefmin}), one finds that 
\be
\phi_{Gen} -\phi_{min} \approx - {1\over 2}\phi_{min} 
+{1\over \sqrt{6} c} (\phi_C-\phi_{min}),
\labeq{figen}
\ee
where the first term dominates. In other words, the fluctuations
we see today were generated at a field value 
approximately half way between the zero and the minimum of
$V(\phi)$.

\section{Cyclic Solutions and Cyclic  Attractors} \label{sec7}

We have shown that a cyclic Universe solution exists provided
we are allowed to pass through the Einstein-frame singularity
according to the
matching conditions elaborated in Section~\ref{sec4},
Eqs.~(\ref{eq:h5in}) and~(\ref{eq:h5mat}). Specifically, we assumed
that $H_5({out}) = -(1+\chi) H_5({in})$ where $\chi$ 
is a non-negative constant, corresponding to branes whose 
relative speed after collision is greater than or equal
to the relative speed
before collision.  Our argument showed that,
for each $\chi\ge 0$, there is a unique  value 
of $H_5({out})$ that is perfectly cyclic.
In the Appendix, we show that an increase in velocity
is perfectly compatible with
energy and momentum conservation in a collision between a 
positive and negative tension brane, provided a greater
density of radiation is generated on the negative tension
brane.  (A similar outcome can occur through the coupling of 
$\phi$ to the matter density, as discussed below Eq.~(\ref{eq:h5mat}),
but we will only discuss the first effect for the purpose of
simplicity.)

In this Section, we wish to show that, under reasonable
assumptions, the cyclic solution is a stable
attractor, typically with a large basin of attraction.  
Without the attractor property, the cyclic model would seem fine-tuned
and unstable. One could imagine that there
would still be brane collisions and periods of contraction and expansion,
but there would be no regularity or long-term predictability to 
the trajectories.
If this were the case, fundamental physics
would lose its  power
 to explain the masses and couplings of elementary particles.
The masses and couplings depend on $\phi$ and other moduli fields.
If there were no attractor solution,
the  precise trajectory of $\phi$
through  cosmic history would depend on initial conditions
and could not be derived from fundamental
physics alone.  In our proposal, the nature of the attractor solution
depends on  
microphysics at the bounce which is computable, in principle,
from fundamental theory.  
Hence, masses and couplings of particles 
change during the course of cycle, but fundamental theory retains
predictive power in determining the way they change and, specifically,
their values at the current  epoch.

The essential feature for attractor behavior
is the extended period of accelerated expansion that 
damps the motion of $\phi$. Let us consider how this works.
Assuming $\chi$ is fixed by microphysics,
there is a value $\bar{H}_5({out})$ which corresponds to the cyclic 
solution. Now, suppose  the  value of $H_5({out})$ exceeds $\bar{H}_5({out})$.
This means that the outgoing velocity exceeds the cyclic value and
$\phi$ runs out farther on the plateau
than in the cyclic case.  Once the field
stops, turns around, and quintessence-domination begins, the field
is critically damped.  By the time $V(\phi)$ falls to zero, the
transient behavior of $\phi$ which depends on the initial value of 
$H_5$ has damped away exponentially so that the field accurately tracks
the slow-roll solution. Following the solution forwards, 
$H_5({in})$ at the next bounce is then exponentially close to 
what it would have been for the cyclic solution. By erasing memory of
the initial conditions, the acceleration insures that $H_5({out})$ after
the next bounce is very nearly  $\bar{H}_5({out})$.

How many e-foldings of accelerated expansion
are actually required to make the cyclic solution an attractor? 
If there is no epoch of accelerated expansion, perturbations will grow
each cycle, becoming self-gravitating and 
non-linear so that no attractor will occur. 
A minimal requirement for obtaining an attractor is that 
linear density perturbations grown during the matter era should
be damped away during the subsequent exponential expansion.
This requires at least 
ln($10^5$)$\sim $
10 e-foldings of exponential expansion. Equally, diluting 
the number density of baryons below one per Hubble volume is 
certainly over-kill in terms of ensuring an attractor, 
and this requires of order 60 e-foldings. In fact, as we discussed
above, obtaining a far larger number of e-foldings is 
perfectly possible.

\begin{figure} 
\begin{center}
\epsfxsize=5.0 in \centerline{\epsfbox{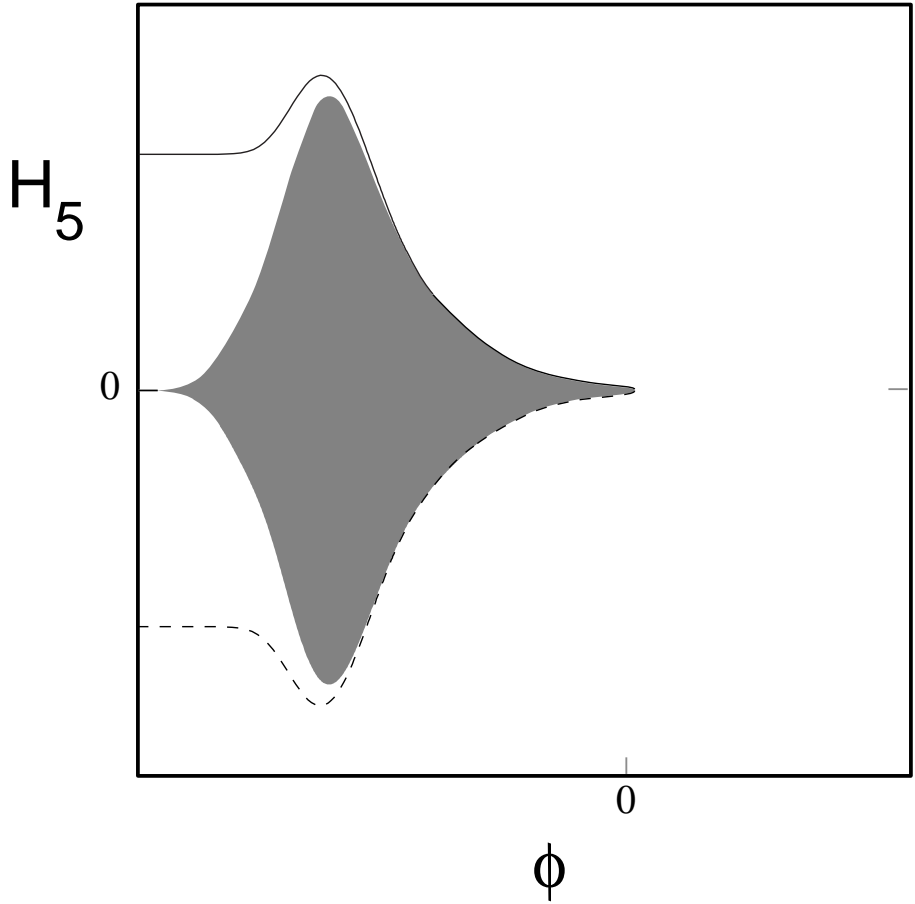}}
\end{center}
\caption{
The cyclic trajectory 
in the $(H_5,\phi)$-plane for the case where no matter and radiation
are produced at the bounce ($\chi=0$).
The grey region
which corresponds to negative energy density, is forbidden.
The solid (dashed) line represents the trajectory
during an expanding (contracting)  phase. Expansion turns to
contraction and vice versa when the trajectory hits the zero
energy surface (the rightmost tip of
grey region in this case).  
     }
\label{phase0}
\end{figure}

\begin{figure} 
\begin{center}
\epsfxsize=5.0 in \centerline{\epsfbox{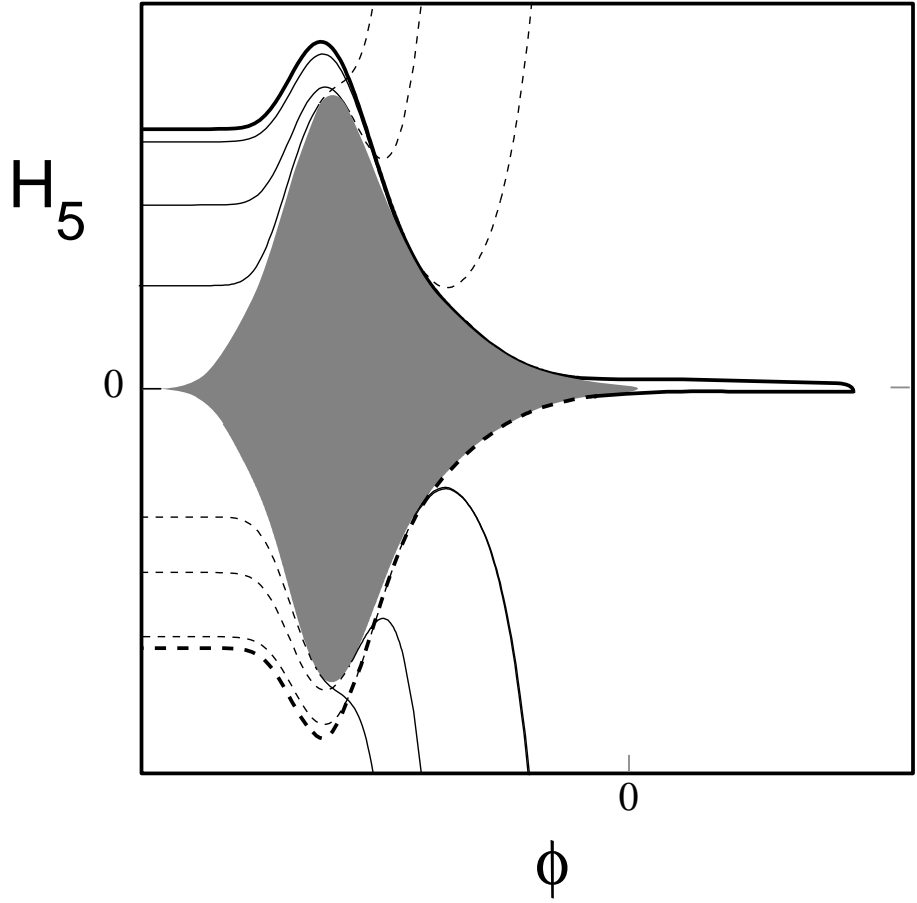}}
\end{center}
\caption{
Trajectories in the $(H_5,\phi)$-plane for the case
where there is radiation.
The solid (dashed) curves represent the trajectory during
an expanding (contracting) phase.
The thin lines illustrate 
undershoot solutions and the heavy line represents an 
overshoot solution.
     }
\label{phase4}
\end{figure}

\begin{figure} 
 \epsfxsize=5.0 in \centerline{\epsfbox{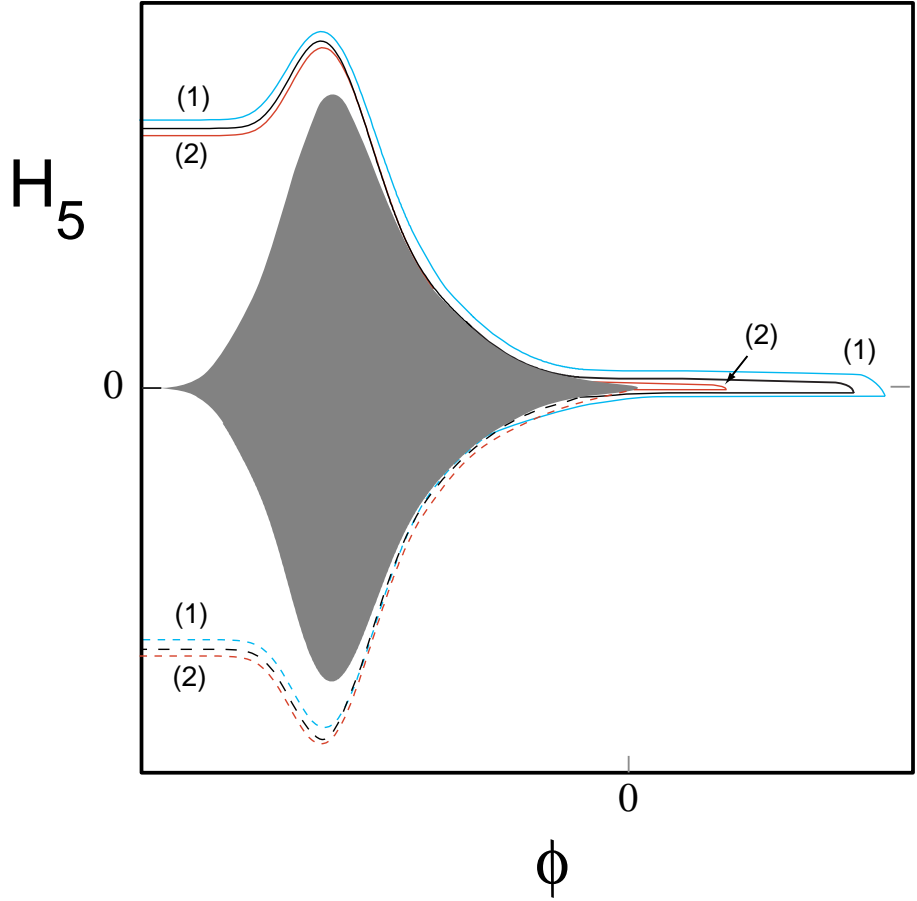}}
	 \caption{
	  Trajectories in the $(H_5,\phi)$-plane showing 
	  the attractor nature of the cyclic solution (the middle
	  trajectory). Path (1) is an overshoot solution
	  that begins with slightly greater
	  velocity ($H_5$) than the attractor, bounces off
	  the grey zero-energy surface, and then has   a contracting
	  trajectory whose value of $H_5$ is smaller in magnitude.
	  Path (2) is an undershoot solution which 
	  begins with slightly less velocity in the
	  expanding phase than the cyclic value and ends in 
	  a contracting phase with $H_5$ having a slightly 
	  greater magnitude. Following the next bounce, therefore,
	   overshoot turns into  undershoot 
	  and vice-versa.   In either case, the deviation 
	  from the attractor value shrinks.
				       }
\label{phasea}
    \end{figure}

To discuss the nature of the attractor solution, it is helpful to
plot the trajectories of the system in the
phase space given by the $(H_5,\phi)$-plane,
shown in Figures~\ref{phase0} through \ref{phasea}.
Recall that $H_5$ is proportional to $\dot{\phi}$;
see Eq.~(\ref{defh5}).
Figure~\ref{phase0} illustrates the cyclic trajectory for the case
where no radiation is generated at the bounce ($\chi=0$) and 
the cycle is exactly time-symmetric.

The phase space plot must always satisfy three properties.
First, for a flat Universe, the Friedmann constraint
equation $H^2={1\over 3} \rho$
implies that the energy density 
$\rho =({1\over 2} \dot{\phi}^2 +V(\phi))$ must be positive.
Without negative space curvature, the system is simply not allowed
to explore negative energies. 
We show the classically excluded region as the shaded area in the
Figures.
In Figure~\ref{phase0} where there is 
no radiation and $V(\phi) \rightarrow 0$
as $\phi \rightarrow 0$, the excluded 
region extends along $H_5=0$  out to $\phi \rightarrow -\infty$.
In Figures~\ref{phase4} and \ref{phasea}
the shape of the  excluded region is modified due to the presence of radiation.
For example, the grey region pinches off on the left hand side for some
finite value of $\phi$.
However, the effect is negligible for the trajectories considered
in our discussion
and so we show the same excluded region in the Figures
as in the case of no radiation.

The second property is that phase space trajectories
are double-valued on the $(H_5,\phi)$-plane.
Given the scalar field and $H_5$, one may have either a
contracting or an expanding Universe. We represent expanding 
trajectories as solid lines and the contracting trajectories
as dashed lines. Two expanding trajectories are not allowed
to cross, and neither are two contracting trajectories
for the usual reasons that hold for particle trajectories 
on phase space. However, an expanding trajectory may certainly
intersect a contracting trajectory. 

The final rule is that there are only two ways 
an expanding trajectory can turn
into a contracting trajectory. If reversal occurs at finite $\phi$
it can only happen if the trajectory hits the forbidden
zero density region (shaded), since $\rho$ has to vanish if
$H$ is to pass smoothly through zero. The shaded region is 
analogous to the ``egg" region described by Brustein and 
Veneziano.\cite{bruV}
The second way in which
contraction can turn into expansion is if the system runs off
to $\phi=-\infty$. Then,  the `bounce' described in Section~\ref{sec3}
and Ref.~\ref{nonsing} occurs. 

The trajectory shown 
in Figure~\ref{phase0} is a cyclic solution (albeit not a 
very interesting one) in which no matter-radiation is produced
at the bounce and the value of 
$H_5({out})$ is precisely equal to $H_5({in})$.  The field 
rolls out in the expanding phase (solid line emanating from the 
upper left side of the Figure) 
to the value where $V(\phi)=0$ and stops (the rightmost 
tip of the grey region).
The total energy density is momentarily zero and expansion reverses to 
contraction. The field then rolls back to $-\infty$ (lower left side
of the Figure).  The 
expanding and contracting phases are exactly symmetrical.

The time-scale
for one cycle of this empty-Universe solution is easily estimated by
noting that 
most of the time is spent
near the zero of the potential, where it is rather shallow,
and the scale factor $a$ is
nearly constant. Therefore, we can neglect
gravity and calculate the period for one cycle in the case of 
the empty Universe:
\be
t_{empty} \approx \int_{\phi_{min}}^0 {d \phi \over 
\sqrt{V_0(e^{-c\phi}-1)}}= {1\over c  \sqrt{V_0}}
\int_0^{c |\phi_{min}|} {dy \over \sqrt{e^y-1}}
\approx {1\over c  \sqrt{V_0}}
\labeq{timec}
\ee
for large $c|\phi_{min}|$. For the parameters typical in the
our examples, this corresponds to roughly
 one
tenth of the current age of the Universe, or a billion 
years.

In Figure~\ref{phase4}, we consider the case
where radiation is produced at the bounce, $\chi >0$.
If $H_5({out})$ is too low compared to the cyclic value,
 the trajectory
encounters the zero density boundary (grey region)
and reverses to contraction.
Solid curves represent the expanding phase of the trajectory, 
and dashed lines represent the contracting phase of the trajectory.
Let us call this an ``undershoot'' solution.
$H_5$ is only constant if $\phi$ and $a$ are both increasing,
or both decreasing. However,  if the Universe reverses when 
$H_5$ is still positive, then  the scalar field kinetic energy
is blue shifted and 
and $H_5$ is rapidly driven to  more and more positive values. 
The trajectory flies off  to large positive $H_5$ and $\phi$
(the upper boundary).

As one increases  $H_5({out})$, the behavior
of the system changes. For sufficiently large $H_5$, the system avoids
the zero energy surface entirely during the period when
$\phi$ is increasing
(the bold solid and dashed trajectory
 in Figure~\ref{phase4}). The field ``overshoots'' the
negative region of the potential and lands on the 
 positive plateau. 
Exponential expansion begins, followed by a very
slow roll of $\phi$ back towards the potential zero,
This period appears as a long, thin excursion on the right hand side
of the figure. $|H_5|$ is small because the field is rolling 
slowing in the quintessence-dominated phase). 

The cyclic attractor solution lies between these undershoot and overshoot
trajectories.
Figure~\ref{phasea} shows trajectories with initial values of $H_5({out})$
both above and below the cyclic value (the middle curve).
Here we can study the stability of the cyclic solution.
Let's first consider
a trajectory with $H_5({out})$ larger than the value in the
cyclic solution. This trajectory is indicated by (1)
in the Figure. Clearly, it overshoots the cyclic 
trajectory and undergoes a longer period of exponential
expansion (the long excursion to the right). 
During the slow-roll epoch, the difference between
this trajectory and the cyclic one damps away until
it is exponentially small. The trajectory encounters
the zero density surface very slightly later than 
the cyclic solution does, and, therefore, reverses  
and ends up with a very slightly smaller value of 
$H_5({out})$  than that in the cyclic trajectory.
Similarly, one can see that starting the system in 
state (2) with a smaller value of $H_5({out})$ than
that of the cyclic trajectory, the system will inflate less and
reverse earlier, ending up with a larger value of 
$H_5({out})$ than that of the cyclic trajectory.
This discussion shows that the trajectory is stable and that memory
of initial decays exponentially after just one cycle.

Here we implicitly assumed that $\chi$ is a constant independent of 
$H_5({in})$, the incoming velocity.  In the Appendix, we obtain
an expression for $\chi$ in Eq.~(\ref{eq:rels}) in terms of the 
matter-radiation
energy densities created on the positive and negative tension branes.
Assuming the energy density on the negative tension brane $\rho_-$ is 
significantly greater than the energy density created on the positive tension
brane, we have $\chi \propto \rho_-$.  
Assuming the collision between branes occurs at a low velocity
so that one is not far from the adiabatic limit, 
$\rho_-$ should decrease with decreasing $H_5{\rm in}$. 
(Note that the low-velocity
assumption  has been made throughout since
it is required for the moduli approximation.) 
Hence, we would anticipate that $\chi$
rises monotonically as
the incoming velocity increases.  
This effect can alter the trajectories and the precise basin of 
attraction in detail, but does not alter the conclusion that a 
large basin of attraction exists. This is assured by having a 
potential plateau or, more generally, a region of the potential in 
which $\phi$ slow-rolls with total energy comparable to the current
dark energy density.

Quantum effects are also unlikely to affect the attractor solution.
We have shown that solution is stable under small perturbations
and here the perturbations remain small since $|V_{min}|$ and 
$V_0$ are small compared to the Planck scale.

What of the trajectories in Figure~\ref{phase4}, for example,
that run away to large $\phi$?  
For these, it is important to understand what happens as $\phi$
grows more positive. One possibility is that the potential $V(\phi)$
diverges as $\phi \rightarrow \infty$.
Our example for $V(\phi)$ has an infinite plateau, but, as discussed
in Section~\ref{sec2}, this is not a general requirement.
If $V(\phi)$ grows sufficiently,
$\phi$ will  bounce back towards $-\infty$.  Alternatively, 
the same effect can occur if
the theory includes massless fields 
that couple to the scale factor
on the negative tension brane, $a_1$. The Lagrangian density 
then includes a term $a_1^4 \dot{\psi}^2  \propto 1/a_1^2$.  
Increasingly positive $\phi$ corresponds to shrinking $a_1$. Hence,
this dynamical term can also create 
a force that causes $\phi$ to bounce back. The net effect
is that $\phi$ rattles back and forth along the potential, possibly
following a chaotic orbit.\cite{damour,moore} These effects
could enhance the basin of attraction for the cyclic solution.
That is,  some of these  
trajectories which we ignored in our undershoot and overshoot
treatment may
 eventually hit the plateau
with low velocity, at which point 
they would  become drawn to the attractor solution.

Finally, let us emphasize that we have only considered
the issue of stability in the context of the
very simplified model 
studied here, with a single scalar field $\phi$,
and the matching conditions discussed in Section~\ref{sec4}.
It would be very interesting
to generalize this discussion to include other moduli,
matter which couples in a nontrivial way to 
$\phi$ as discussed in Section~\ref{sec4}, and 
and also discrete degrees of freedom such as
a quantized four-form field, which may change from cycle
to cycle so that the system really explores moduli space.
The existence or otherwise of an attractor could well
be relevant to the determination of the relative abundances
of dark energy, dark matter, baryons and photons in
the Universe, and also to the values of the fundamental constants
of nature.

\section{Implications}  \label{sec8}

The strengths of the cyclic model are its simplicity, its 
efficient use of all of the dominant elements of the Universe and 
the fact that it is a complete description of all phases of 
cosmic evolution.   
This can be contrasted with inflationary
cosmology, a highly appealing theoretical model in its own right.  
Inflationary
cosmology focuses on a brief epoch when the Universe was 
$10^{-35}$~seconds old.  The model relies on assumptions about 
how the Universe emerged from the cosmic singularity.  
One must postulate the existence of a phase of rapid cosmic
acceleration at very high energies, for which their exists no
direct proof.   (In contrast, the cyclic model relies on 
low energy cosmic acceleration that has been observed.)
Subsequent cosmic events, such as the recent transition from 
matter-domination to dark energy domination and cosmic 
acceleration, appear to have no direct 
connection to inflationary theory.

Because the cyclic model ties the past, present and future evolution
of the Universe in a tight, cross-correlated way, it
 has surprising explanatory and 
predictive power.   In the introduction to this paper, we noted
a number of the most challenging questions of cosmology and 
fundamental physics.  In this section, we consider each of these
questions (and more) and briefly describe
the insights the cyclic model provides concerning 
their answers.  

\subsection{Why is the Universe homogeneous, isotropic and flat?}

The Universe is made homogeneous and isotropic during  the 
period of the preceding  cycle when quintessence dominates and 
the Universe is 
undergoing slow cosmic acceleration.  This ensures that the 
branes are flat and parallel as they begin to approach, collide,
and emerge in a big bang.   Inflation also relies on cosmic 
acceleration, but driven by very high vacuum energy which
produces an acceleration that is nearly
$10^{100}$ times faster.

\subsection{How were density  inhomogeneities  generated?}

In the cyclic model,
the observed inhomogeneities in the Universe are generated during 
the contracting phase  when the scale factor is nearly static
and gravitational effects are weak. Consequently, as in the 
ekpyrotic scenario, a nearly scale invariant 
spectrum of adiabatic, gaussian
energy density fluctuations is generated. However,
because the expansion rate is negligible and gravitational 
effects are weak,
the tensor (metric fluctuation) spectrum is blue with an exponentially
tiny amplitude at long wavelengths.  

Fluctuations are also created during the quintessence dominated phase,
just as they are during inflation.  However, because the energy 
density during the accelerating phase is 100 orders of magnitude 
smaller than in inflation, the resulting fluctuation amplitude is 
exponentially smaller in the cyclic model.  These fluctuations also
have wavelengths that exceed the current Hubble horizon.  Hence,
they are observationally irrelevant.

\subsection{What is the role of dark energy and the current 
cosmic acceleration?}

Clearly, dark energy and the current cosmic acceleration play
an essential role in the cyclic model both by reducing the 
entropy and black hole density of the previous cycle, and
triggering  the turnaround  from an expanding to a contracting phase.
(In all other cosmologies to date, including inflation, dark 
energy  has no essential role.)

\subsection{How old is the Universe?}

A truly cyclic Universe is clearly infinitely old in terms
of cosmic time. As we have noted, the 
exact cyclic solution can also be an attractor. Hence, the cycling is 
stable.  Consequently, one becomes insensitive to
the initial conditions for the Universe as long as 
they were within the basin of attraction of the cyclic
solution. We believe that within this framework, the
problem of the initial conditions for the universe is
significantly altered: as long as the universe has some
nonzero probability for entering the cyclic solution,
large regions of the universe maintain cyclic evolution
 for  arbitrarily long periods of time. 

There is a possible objection to this argument, due to
the fact that the four dimensional nonsingular brane 
spacetimes in our scenario are past geodesically incomplete.
As we have explained, for most of cosmic time
they are well approximated by de Sitter
spacetime, with a cosmological constant (or vacuum energy)
close to the currently observed value. This nearly de Sitter 
spacetime is foliated by slices of constant scalar field
$\phi$, which are nearly geometrically flat.
Matter is 
repeatedly generated on the slices with $\phi=-\infty$,
in the rest frame defined by those slices.  

As one follows cosmic time $t$ backwards, one must pass
an infinite number of these big crunch-big bang surfaces.
However, even though the cosmic time tends to $-\infty$, the
proper time as measured along timelike geodesics running into the
past generically is finite even as $t$ tends to $-\infty$. 
This may be seen as follows. Consider a particle with 
momentum $P$ in the flat slicing. Its momentum blueshifts as
$a^{-1}$ as you follow it back in time. 
The geodesic becomes nearly null and the 
proper time measured along the geodesic converges even though
$t$ tends to $-\infty$ (this is the crux of the recent
argument of Borde, Guth and Vilenkin that inflation
is past geodesically incomplete \cite{BGV}).

In our scenario, however, {\it all physical particles 
are created with finite momentum in the flat slicing defined
by $\phi$}. If we follow a particle present in today's 
universe back in time, most likely it was created 
on the last $\phi=-\infty$ surface. With an exponentially
smaller probability, it could have been created 
on the penultimate $\phi=-\infty$ surface, and
so on into the past. The probability that any
observed particle originated on the $t=-\infty$
flat surface which is the boundary of the flat
slicing of de Sitter spacetime, is zero. 
Therefore we do not attribute any physical significance
to the past geodesic incompleteness of the spacetime
metric in our scenario.
In contrast, particle production in 
standard inflationary models 
occurs  on open slices  a finite time ago.

Even if there are no particles present which `saw' the past 
boundary of the cyclic universe, one might object that
the scenario implicitly requires a boundary condition in the infinite
past. We do not think this is a strong objection. 
If the cyclic solution were
begun within a finite region (for example a torus) of
three dimensional space, it would grow exponentially
with each cycle to an arbitrarily large size. 
After an arbitrarily long time, to any real observer
the universe would appear to be infinite both in
spatial extent, and in lifetime to the past.

So, whilst the cyclic model
still requires an
initial condition, provided that state is within the 
basin of attraction of the cyclic solution, 
we are completely insensitive to its details. Any features
of the initial state (the total size of the Universe,
or any fluctuations about flatness or homogeneity), become
exponentially diluted in each cycle and since the 
cyclic solution can repeat forever, are ultimately 
completely irrelevant to any observation.

\subsection{What is the  ultimate fate of the Universe?}

The cycles can be continued to the infinite future, as well as the
infinite past.   
Hence, the Universe endures forever.

\subsection{How big is the Universe?}

From the effective 4d point-of-view, the Universe oscillates between
periods of expansion to periods of contraction down to a big crunch.
However, from the brane world point-of-view, the Universe is always
infinite in the sense that the branes always have infinite extent.
The fact that the branes are spatially infinite means that it 
is possibly for the total entropy in the Universe to increase
from cycle to cycle, and, at the same time, have the entropy density
(in particular,  the total
entropy per Hubble horizon) become nearly zero prior to 
each bounce.

\subsection{What occurs at the big bang singularity?}

The cyclic model utilizes the  ekpyrotic notion
that the singularity corresponds to the collision and bounce of
two outer orbifold branes in a manner that is continuous and
well-behaved. The singularity is not a place where energy and 
curvature diverge and time begins. Rather, formulated in appropriate
fields and coordinates, the singularity is a smooth, finite
transition
from a contracting phase heading towards a big crunch and 
a big bang evolving into an expanding Universe.

\subsection{What determines the arrow of time?}

Since the Universe is cyclic, it may appear that there is no well-defined
means of determining the arrow of time.  Indeed, for a local observer,
there is no clear means of doing so. 

From the global perspective, though, 
there is a clear means of determining
forward from backward in time.  
First, one of
the boundary branes is forever expanding
in the ``forward" time direction 
in the cyclic model. The other brane is expanding except for 
brief intervals of contraction, but, averaged over a cycle, the 
net effect is expansion.
The rate changes from phase to phase, as well 
as the separation.   In the contraction phase, the branes themselves
stretch at a rate that  is 
slow but their separation rapidly  decreases. In the radiation, matter,
and quintessence dominated phases, the branes stretch significantly, 
but
their separation remains fixed.  During this period, the entropy 
created during the previous cycle is spread out exponentially, reducing 
the degrees of freedom per horizon to nearly zero.

\subsection{Why is the cosmological constant so small?}

The cyclic model provides a fascinating new outlook on this vexing 
problem.  Historically, the problem is assumed to mean that one must
explain why the vacuum energy of the ground state is zero.  

In the cyclic model, the vacuum energy  of the ground state is 
not zero.  It is negative and its magnitude is large, as is obvious
from Figure~1.
If the Universe begins in the ground state, the negative cosmological
constant will cause  rapid recollapse, as expected for an
anti-de Sitter phase. In the cyclic scenario, though, 
we have shown how to arrange
conditions where the Universe avoids the ground state.
Instead, the Universe 
hovers from cycle to cycle  above the ground state
bouncing from one side of the potential well
to the other but spending most time on the positive energy side.
The branes are moving too rapidly whenever  the separation 
corresponds to the potential minimum.

There remains the important challenge of explaining why the 
the current potential energy is so small.  The value depends on 
both the shape of the potential curve and the precise transfer of
energy and momentum
at the bounce.  Perhaps explaining the value will be an
issue as knotty as the cosmological constant problem, or perhaps
the conditions will prove easier to satisfy. 
What is certain, though,
is that the problem is shifted from tuning a vacuum energy, 
and this provides an opportunity for new kinds of solutions.

\subsection{Equation-of-state of dark energy}

The equation-of-state of the dark energy, $w$, is the ratio 
of the pressure to the energy density of $\phi$,
$(\frac{1}{2} \dot{\phi}^2 -V)/ (\frac{1}{2} \dot{\phi}^2 +V)$.
In Section~\ref{sec5}, we discussed the evolution of $\phi$ in 
the radiation, matter  and quintessence  dominated epochs. 
The generic result is that evolution of $\phi$ in the positive
direction halts and the field begins to roll back towards $-\infty$
in the radiation-dominated epoch.  At the turn-around, $w=-1$
since the kinetic energy is zero.  As the field rolls back and its
kinetic energy increases, $w$ increases.  Hence, the 
generic result is that $w$ is close to -1 today and increasing.
Conceivably, cosmological observations could detect this prediction.
Tracker models of quintessence, some of the best-motivated
alternatives, have the opposite trend: $w$ 
is near $-0.8$ or so today and decreasing towards $-1$.\cite{Zlatev}
Other models, such as $k$-essence, have the same trend as found
in the cyclic model.\cite{kess}  

\subsection{Implications for Supersymmetry and Superstrings}

The cyclic model imposes different constraints 
on fundamental physics compared to previous cosmological models.
As an example,
consider the problem of  designing supergravity
potentials.  The potentials   are constructed from a 
superpotential $W$ according to the prescription:
\begin{equation}
V=e^{K/M_{pl}^{2}}\left[K^{ij}D_i W \,  \bar{D_j W}-\frac{3}{M_{pl}^{2}}
W\bar{W}\right].
\end{equation}
where $D_i = \partial/\partial \phi^i + K_i/M_{pl}^{2}$ is the K\"ahler covariant derivative,
$K_i = \partial K/ \partial \phi^i$,
$K_{ij} = \partial^2 K/ \partial\phi^i \partial\phi^j$ and a sum over each
superfield $\phi_i$ is implicit.
If the ground state is supersymmetric, $D_i W =0$, the first term
is zero.  In general, unless $W$ is zero  for precisely the
same values for which 
$D_i W =0$, 
the minimum has a negative cosmological constant. In the past, this 
type of model would have been ruled unacceptable.  The possibility of
a cosmology in which the Universe hovers over the ground state in a 
state of zero or positive energy revives these models and  alters
constraints on model building.  

An obvious  but important implication
is that  supersymmetry breaking can be achieved without having
 spontaneous symmetry breaking in the ground state. In this scenario,
 it suffices if the  Universe hovers in the radiation, matter and 
 quintessence dominated epochs at some state far above the 
 ground state in energy and that the supersymmetry is broken 
 by the appropriate amount in the
 hovering state, where the radiation, matter and quintessence
 dominated phases occur.
These considerations have a significant impact on the design of 
phenomenological  supersymmetric models.

One other requirement/prediction of the cyclic scenario (and the
ekpyrotic models in general)
is  that the branes move
in a space-time with  co-dimension one.  
The constraint derives from having a bounce that produces a smooth
transition from contraction to expansion.
As argued by Khoury {\it et al.}, 
the geometry is flat arbitrarily close to the bounce provided 
there is one extra dimension only. Hence, brane world scenarios
based on theories like that of Ho\v rava and Witten are acceptable,
but large extra-dimensional models relying on having co-dimension
two or greater are problematic.

\subsection{Hoyle's Revenge?}

Within  each cycle, 
there is  a sequence of kinetic energy, radiation, 
matter and quintessence dominated phases of 
evolution that are in accord with the standard big bang cosmology.
However, averaged over many cycles, the model can be viewed as 
a remarkable  re-incarnation of Fred Hoyle's 
steady state model of the Universe.
Most of the cycle is spent in a  phase with nearly
constant energy density, as in the steady state picture.
Indeed 
Hoyle's $C$-field that was introduced to provide a 
constant supply of 
matter (and a preferred rest frame)
is replaced by our scalar field $\phi$, which 
defines a preferred time slicing and 
generates matter repeatedly at each bounce, 
restoring the Universe 
to a state of high temperature and matter density.
In Hoyle's steady state model, every flat spatial slice was
statistically identical.  Here the slices are identical only when
separated by one period, so we have a discrete rather than continuous
time translation symmetry. Nevertheless when coarse grained over 
large time spans, the structure is similar to that proposed
in the steady state Universe. 
Global properties of the cyclic cosmology will be discussed
elsewhere.\cite{AGST}

\vspace*{.1in}
\noindent
{\bf Acknowledgements:}
We thank M. Bucher, S. Gratton, D. Gross, A.Guth,
J. Khoury, B.A. Ovrut, J. Ostriker, P.J.E. Peebles, A. Polyakov, M. Rees,
N. Seiberg, D. Spergel, A. Tolley, A. Vilenkin, 
T. Wiseman and E. Witten for useful
conversations.
We thank L. Rocher for pointing out Ref.~\ref{rocher}
and other historical references.
    This work was supported in part by
      US Department of Energy grant
      DE-FG02-91ER40671 (PJS) and by PPARC-UK (NT).

\section{Appendix: Matching $H_5$ across the Bounce}

In this appendix we discuss the matching condition needed to
determine $H_5({out})$ in terms 
of $H_5({in})$. We shall assume that 
all other extra dimensions and moduli are fixed, and the bulk space-time
between the branes settles down to a static 
state after the collision.
(In the simplest brane world models, there is a Birkhoff theorem
which ensures that there is a coordinate system in which 
the bulk metric is static in between the branes).
We shall take the densities of radiation on the
branes after collision as being given. 
By imposing 
Israel matching in both initial and final states, 
as well as conservation of total energy and momentum, 
we shall be able to completely fix the state of the
outgoing branes and in particular the expansion rate of
the extra dimension
$H_5({out})$, in terms of $H_5({in})$.
 A more complete discussion of this
method will be presented in 
Ref.~\ref{martinneil}.

The idea is to treat the brane collision as a short-distance phenomenon.
The warp factor may be 
treated as linear 
between the branes as they approach or recede. 
Linearity plus  $Z_2$ symmetry 
ensures that
the kinks in the warp factors are 
equal in magnitude and opposite in sign.
Israel matching relates the kink magnitudes to
the densities and 
speeds of the branes, yielding the 
relations we use below.

The initial state of empty branes with tensions $T$ and $-T$, and
with corresponding velocities $v_+<0$ and $v_->0$ (measured
in the frame in which the bulk is static) obeys
\ba
T\sqrt{1-v_+^2} &=&
T \sqrt{1-v_-^2}\cr
E_{tot} &=& 
{T\over 
\sqrt{1-v_+^2}} - 
{T\over 
\sqrt{1-v_-^2}}\cr
P_{tot} &=& 
{T v_+\over 
\sqrt{1-v_+^2}} - 
{Tv_-\over 
\sqrt{1-v_-^2}}.
\labeq{initthree}
\ea
The first equation follows from Israel matching on 
the two branes as the approach, and equating the
kinks in the brane scale factors.
The second and third equations are the definitions of the total
energy and momentum.
The three  equations (\ref{eq:initthree}) 
imply that the incoming, empty state has
$v_+=-v_-$, $E_{tot}=0$ and that the
total momentum is
\be
P_{tot} = {TL H_5({in}) \over \sqrt{1-{1\over 4} (LH_5({in}))^2}} <0,
\labeq{ptot}
\ee
where we identify $v_+-v_-$ with the contraction speed of
the fifth dimension, $|LH_5({in})|$.

The corresponding equations for the outgoing state 
are easily obtained, by replacing $T$ with
$T+\rho_+\equiv T_+$ for the positive tension 
brane, and $-T$ with $-T+\rho_-\equiv -T_-$ for the negative tension
brane, assuming the densities of radiation produced at the collision
on each brane, $\rho_+$ and $\rho_-$ respectively, are given
from a microphysical calculation, and are both positive.

We now wish to apply energy and momentum conservation, and
Israel matching to the final state. 
The only subtlety is that 
the $(t,y)$ frame in which the bulk is static is not necessarily the
same frame in the final state as it was in the initial state,
so one should boost the initial two-momentum $(E_{tot},P_{tot})$
with a velocity $V$ and then apply the Israel constraints 
and energy-momentum conservation equations in the new boosted frame.
The latter provide three equations for the three unknowns in the
final state, namely $v_+(out) $, $v_-(out) $ 
and $V$. Writing $v_\pm(out) = {\rm tanh}(\theta_\pm)$, where
$\theta_\pm$ are the associated rapidities, one obtains 
two solutions 
\ba 
{\rm sinh} \theta_+ &=& 
-{1\over 2 T_-} \left(|P_{tot}|+ |P_{tot}|^{-1}(T_+^2-T_-^2)\right)\cr
{\rm sinh} \theta_- &=& 
\pm{1\over 2 T_+}
\left(|P_{tot}|- |P_{tot}|^{-1}(T_+^2-T_-^2)\right),
\labeq{arraysin}
\ea
where $T_+\equiv T+\rho_+$, $T_-\equiv T-\rho_-$ with $\rho_+$ and $\rho_-$
the densities of radiation 
on the positive and negative tension branes
respectively, after collision. Both $\rho_+$ and $\rho_-$
are assumed to be positive.
In the first solution, with signs $(-+)$, the velocities of
the positive and negative tension branes are the same 
after the collision as they were before it. In the second,
with signs $(--)$, 
the positive tension brane continues in the negative $y$ 
direction but the negative tension brane is {\it also}
moving in the negative $y$ direction. 

The corresponding values for $v_\pm(out)$ and $V$ are
\ba
v_+({out})&=& 
-{|P_{tot}|+ |P_{tot}|^{-1}(T_+^2-T_-^2) \over \sqrt{ P_{tot}^2 +2(T_+^2+T_-^2) +P_{tot}^{-2} (T_+^2-T_-^2)^2}},\cr
v_-({out})&=& 
\pm{|P_{tot}|- |P_{tot}|^{-1}(T_+^2-T_-^2) \over \sqrt{ P_{tot}^2 +2(T_+^2+T_-^2) +P_{tot}^{-2} (T_+^2-T_-^2)^2}},\cr
V&=& - {\sqrt{ P_{tot}^2 +2(T_+^2+T_-^2) +P_{tot}^{-2} (T_+^2-T_-^2)^2}\over
|P_{tot}|(T_+^2+T_-^2)/(T_+^2-T_-^2)+|P_{tot}|^{-1} (T_+^2-T_-^2)},\cr
{\rm or}\quad &=&  - {\sqrt{ P_{tot}^2 +2(T_+^2+T_-^2) +P_{tot}^{-2} (T_+^2-T_-^2)^2}
\over |P_{tot}|+|P_{tot}|^{-1}(T_+^2+T_-^2)},
\labeq{solthree}
\ea
where the first solution for $V$ holds for the $(-+)$ case,
and the second for the $(--)$ case.

We are interested in the
relative speed of the branes in the outgoing state, since
that gives the expansion rate of the extra dimension,
$-v_+({out})+v_-({out})=L H_5({out})$, compared to their
relative speed $-2 v_+= -L H_5({in})$ in the incoming state. We find
in the $(-+)$ solution,
\be
\left|{H_5{(out})\over H_5({in})}\right| 
= {v_+({out})-v_-({out})\over 2 v_+}
= \sqrt{P_{tot}^2+4 T^2\over 
P_{tot}^2 +2(T_+^2+T_-^2) +P_{tot}^{-2} (T_+^2-T_-^2)^2},
\labeq{rels}
\ee
and in the $(--)$ solution 
\be
\left|{H_5{(out})\over H_5({in})}\right| 
= {(T_+^2-T_-^2)\over P_{tot}^2}  \sqrt{P_{tot}^2+4 T^2\over 
P_{tot}^2 +2(T_+^2+T_-^2) +P_{tot}^{-2} (T_+^2-T_-^2)^2}.
\labeq{relsa}
\ee
with $P_{tot}$ given by (\ref{eq:ptot}) in both cases.

At this point we need to consider how the densities
of radiation $\rho_+$ and $\rho_-$ depend on the 
relative speed of approach of the branes. At very low speeds,
$|LH_5(in)| <<1$, 
one expects the outer brane collision to be nearly adiabatic
and an exponentially small amount of radiation to be produced.
The $(-+)$ solution has the speeds of both branes 
nearly equal before and after collision: we assume that
it is this solution, rather than the $(--)$ solution
which is realised in this low velocity limit.

As $|LH_5(in)|$ is increased, we expect $\rho_+$ and
$\rho_-$ to grow. Now, if we consider $\rho_+$ and $\rho_-$ to be both 
$<<P_{tot}<<T$, 
then the second term in the denominator
dominates. If more radiation is produced on the negative tension
brane, $\rho_->\rho_+$, then $|H_5({out})/H_5({in}) |\equiv
(1+\chi) \approx
(1+ (\rho_--\rho_+)/2T)$ and so $\chi$ is small and positive.
This is the condition noted in the text, necessary to obtain 
cyclic behavior. 
Conceivably, the brane tension can change from $T$ to $T'= T-t$ at
collision.  Then, we obtain $(1+\chi) \approx
(1+ (\rho_--\rho_+  + 2 t)/2T)$

For the $(-+)$ solution, we 
can straightforwardly determine an upper limit
for $|H_5({out})/H_5({in}) |\equiv (1+\chi)$.  Consider, for example,
the case there the brane tension is unchanged at collision, $t=0$.
The expression
in 
(\ref{eq:rels}) gives
 $|H_5({out})/H_5({in}) |$ as a function of $T_+$, $T_-$ and
$P_{tot}$. It is greatest, at fixed $T_-$ and $P_{tot}$, 
when $T_+=T$, 
its smallest value. For $P_{tot}^2<T^2$, it is maximized 
for $T_-^2=T^2-P_{tot}^2$, and  equal to
$\sqrt{1+P_{tot}^2/(4T^2)}$ when equality holds. 
For $P_{tot}\geq T^2$, it is maximized when $T_-=0$,
its smallest value, and $P_{tot}^2=2T^2$, when it
is equal to $\sqrt{4\over 3}$.
This is more than enough for us to obtain the small
values of $\chi$ needed to make the cyclic scenario work.
A reduction in brane tension at collions $t>0$ further increases the
maximal value of the ratio.  To obtain cyclic behavior, we need $\chi$
to be constant from bounce to bounce.
That is, compared to the tension before collision,
the fractional change in tension and the fractional
production of radiation must be constant.

We shall not consider the $(--)$ solution in detail,
except to note that in the small $P_{tot}$ limit
it allows an arbitrarily large value for
$|H_5{(out})/H_5({in})|$, which seems unphysical.

Let us re-iterate that,
there are many caveats  attached to this calculation. We
have not calculated $\rho_+$ and $\rho_-$ and have left these
as parameters. We have set the brane tensions $T$ to be
equal before and after the collision, and have neglected
possible bulk excitations, treating the bulk spacetime as static 
both before and after collision. We have ignored matter couplings to 
bulk scalars, and ignored the possible dynamical evolution of
additional extra dimensions. Nevertheless we think it encouraging
that the unusual behavior of matter bound to a negative tension
brane allows $H_5({out})/H_5({in})$ to be slightly greater than
unity, which is
what we need for cyclic behavior.


\begin{thebibliography}{9999}
\bibitem{ST1} P.J. Steinhardt and N. Turok, hep-th/0111030.
\bibitem{Rocher} See, for example, J. Hastings, {\it Encyclopaedia
of religion and ethics}, (C. Scribner \& sons, New York, 1927).
\label{rocher}
\bibitem{Tolman} R.C. Tolman, {\it Relativity,  Thermodynamics and
Cosmology}, (Oxford U. Press, Clarendon Press, 1934).
\bibitem{DP} See, for example, R.H. Dicke and J.P.E. Peebles,
in {\it General Relativity: An Einstein Centenary Survey},
ed. by S.H. Hawking and W. Israel, (Cambridge U. Press, Cambridge,
1979).
\bibitem{PBB} G. Veneziano, {\it Phys. Lett. B}{\bf 265}, 287  (1991);
 M. Gasperini and G. Veneziano, Astropart. Phys. {\bf 1},
(1993) 317.
\bibitem{rev}
For a review, see  N. Bahcall, J.P. Ostriker. S.
 Perlmutter, and 
 P.J. Steinhardt, {\it Science} {\bf 284}, 1481-1488, (1999).
\bibitem{kost1} J. Khoury, B.A. Ovrut,
P.J. Steinhardt and N. Turok, hep-th/0103239, {\it Phys. Rev. D}, in press.
\label{kost1}
 \bibitem{nonsing}
J. Khoury, B.A. Ovrut,
 N. Seiberg,
  P.J. Steinhardt and N. Turok,
  hep-th/0108187.
  \label{nonsing}
  \bibitem{SP} N. Seiberg and J. Polchinski, private communication.
  \label{SP}
\bibitem{PGT} J. Garriga, E. Pujolas and T. 
Tanaka hep-th/0111277.
\label{PGT}
\bibitem{SST} N. Seiberg, P.J. Steinhardt and N. Turok, in preparation.
\bibitem{ekperts}
  J. Khoury, B.A. Ovrut,
P.J. Steinhardt and N. Turok, hep-th/0109050.
\label{ekperts}
\bibitem{sw} J. Feng, J. March-Russell, S. Sethi, and F. Wilczek, 
{\it Nucl. Phys. B}{\bf 602}, 307 (2001). 
\bibitem{bp} R.Bousso and J. Polchinski,
{\it JHEP} {\bf 0006}, 006 (2000).
\bibitem{quint}  R.R. Caldwell, R. Dave and P.J. Steinhardt, {\it Phys.
Rev. Lett.} {\bf 80}, 1582 (1998); see also, J. Ostriker and
P.J. Steinhardt, {\it Sci. Am.}, January, 2001.
\bibitem{fut} J. Khoury, B. Ovrut, P.J. Steinhardt and N. Turok,
in preparation.
\bibitem{fock} V. Fock, {\it Theory of Space, Time and Gravitation},
Pergamon, London, 1959. 
\bibitem{other}
N. Turok and P.J. Steinhardt, in preparation.
\bibitem{poly2} T. Damour and A. Polyakov,
{\it Nucl. Phys. B}{\bf 423}, 532 (1994).
\bibitem{poly}
T. Damour, gr-qc/0109063.
\bibitem{bruV} R. Brustein and G. Veneziano,
{\it Phys. Lett. B}{\bf 329}, 429 (1994);
R. Brustein and R. Madden, {\it Phys. Lett. B}{\bf 410}, 110 (1997).
\bibitem{damour} T. Damour and M. Henneaux,
{\it Phys. Rev. Lett.} {\bf 85}, 920 (2000).
\bibitem{moore} J.H. Horne and G. Moore, {\it Nucl. Phys. B}{\bf 432},
109 (1994).
\bibitem{BGV} A. Borde, A. Guth and A. Vilenkin,
gr-qc/0110012.
\bibitem{Zlatev}  I. Zlatev and P.J. Steinhardt, {\it Phys. Lett. B}{\bf 459},
570 (1999);
I. Zlatev, L. Wang and P.J. Steinhardt,
{\it Phys. Rev. D}{\bf 59}, 123504 (1999).
\bibitem{kess}
  C. Armendariz-Picon, V. Mukhanov, and P.J. Steinhardt,
  {\it Phys. Rev. Lett.} {\bf 85}, 4438  (2000).
\bibitem{AGST} A. Aguirre, S. Gratton, P.J. Steinhardt,
N. Turok, to appear.
\bibitem{martinneil} M. Bucher and N. Turok, in preparation.
\label{martinneil}
\end{thebibliography}
\end{document}